\documentclass[10pt,journal,compsoc]{IEEEtran}
\usepackage{comment}

\usepackage{multirow}

\usepackage{paralist} 

\usepackage{float}

\usepackage{algorithm}
\usepackage{algpseudocode}

\usepackage{amsmath}

\usepackage{hyperref}

\usepackage{url}

\usepackage{enumitem}

\usepackage{booktabs} 

\usepackage{xcolor}

\usepackage{bbm}

\ifCLASSOPTIONcompsoc
  \usepackage[nocompress]{cite}
\else
  \usepackage{cite}
\fi
\ifCLASSINFOpdf
  \usepackage[pdftex]{graphicx}
  \graphicspath{{../pdf/}{../jpeg/}}
  \DeclareGraphicsExtensions{.pdf,.jpeg,.png}
\else
  \usepackage[dvips]{graphicx}
  \graphicspath{{../eps/}}
  \DeclareGraphicsExtensions{.eps}
\fi
\newcommand{\rc}[1]{\textcolor{black}{#1}}

\begin{document}
%
\title{FraudAuditor: A Visual Analytics Approach for Collusive Fraud in Health Insurance}
%
%
%
%

\author{Jiehui~Zhou,
        Xumeng~Wang,
        Jie~Wang,
        Hui~Ye,
        Huanliang~Wang,
        Zihan~Zhou,
        Dongming~Han,
        Haochao~Ying,
        Jian~Wu,
        and~Wei~Chen
\IEEEcompsocitemizethanks{\IEEEcompsocthanksitem J. Zhou, H. Wang, Z. Zhou, D. Han, and W. Chen are with the State Key Lab of CAD\&CG, Zhejiang University.
E-mail: \{zhoujiehui, wanghuanliang, zhouzihan, dongminghan, chenvis\}@zju.edu.cn.
\IEEEcompsocthanksitem X. Wang is with TMCC, CS, Nankai University.
E-mail: wangxumeng@nankai.edu.cn.
\IEEEcompsocthanksitem J. Wang is with Alibaba Group, Hangzhou.
E-mail: siwei.wj@alibaba-inc.com.
\IEEEcompsocthanksitem H. Ye is with Tencent, Shenzhen.
E-mail: hazelye@tencent.com.
\IEEEcompsocthanksitem H. Ying is with the School of Public Health, Zhejiang University. He is also with the Key Laboratory of Intelligent Preventive Medicine of Zhejiang Province.
E-mail: haochaoying@zju.edu.cn.
\IEEEcompsocthanksitem J. Wu is with Second Affiliated Hospital School of Medicine, School of Public Health, and Institute of Wenzhou, Zhejiang University.
E-mail: wujian2000@zju.edu.cn.
\IEEEcompsocthanksitem Wei Chen and Haochao Ying are the corresponding authors.
}
\thanks{Manuscript received xx xx, 20xx; revised xx xx, 20xx.}}

%
%

\markboth{Journal of \LaTeX\ Class Files,~Vol.~xx, No.~x, xx~20xx}%
{Shell \MakeLowercase{\textit{et al.}}: Bare Demo of IEEEtran.cls for Computer Society Journals}
%



\IEEEtitleabstractindextext{%
\begin{abstract}
Collusive fraud, in which multiple fraudsters collude to defraud health insurance funds, threatens the operation of the healthcare system. 
However, existing statistical and machine learning-based methods have limited ability to detect fraud in the scenario of health insurance due to the high similarity of fraudulent behaviors to normal medical visits and the lack of labeled data.
To ensure the accuracy of the detection results, expert knowledge needs to be integrated with the fraud detection process. By working closely with health insurance audit experts, we propose \textit{FraudAuditor}, a three-stage visual analytics approach to collusive fraud detection in health insurance. Specifically, we first allow users to interactively construct a co-visit network to holistically model the visit relationships of different patients. Second, an improved community detection algorithm that considers the strength of fraud likelihood is designed to detect suspicious fraudulent groups. Finally, through our visual interface, users can compare, investigate, and verify suspicious patient behavior with tailored visualizations that support different time scales. We conducted case studies in a real-world healthcare scenario, i.e., to help locate the actual fraud group and exclude the false positive group. The results and expert feedback proved the effectiveness and usability of the approach.
\end{abstract}

}

\maketitle

\IEEEdisplaynontitleabstractindextext

%
\IEEEpeerreviewmaketitle

\IEEEraisesectionheading{\section{Introduction}\label{sec:introduction}}

\IEEEPARstart{A}{n} effective health insurance system plays a significant role in managing healthcare resources, enhancing life quality for people, and maintaining social stability. More than 1.3 billion people have enrolled in the National Basic Medical Insurance in China\footnote{\url{http://en.nhc.gov.cn/2020-06/28/c\_80923.htm}}. However, increasing health insurance fraud events have become a severe social problem. According to the inspection conducted by the National Healthcare Security Administration and the Ministry of Public Security in China, nearly half of 815,000 health institutes have improper or even illegal funds costs in 2020, leading to an economic loss of more than 22.3 billion yuan (\$3.4 billion)\footnote{\url{http://en.ce.cn/main/latest/202102/22/t20210222\_36327961.shtml}}.
The emerging collusive fraud is the most serious and urgent among these events~\cite{li2008survey}. Fraudsters collude to purchase drugs with insurance reimbursement and cash the drugs out.
The massive amount of fraud brings serious consequences. There is an urgent need for efficient and effective detection methods to quickly identify collusive fraud and prevent further loss.

Detecting collusive fraud in health insurance faces two challenges. First, it is difficult to distinguish the medical visits behavior of fraudsters from those of normal patients.
Typically, fraudsters frequently buy large quantities of easily marketable drugs.
However, due to the need to maintain long-term medication, patients with chronic diseases and those requiring Traditional Chinese medicine (TCM) treatment have similar purchasing behaviors to fraudsters'.
Second, manual auditing is necessary but laborious.
Misidentification is unacceptable for fraud detection because a patient has to bear legal responsibility after being recognized as a fraudster. Verifying fraud requires auditors to synthesize a large amount of contextual information, such as the amount of reimbursement, the degree to which the patient's disease and drugs match, and the time of visits.

Existing collusive fraud detection methods can hardly handle these challenges. Existing methods focus on modeling the relationships between fraudsters by graphs and detecting fraudulent groups through statistical or machine learning (ML) methods. Statistical approaches use structural and attribute features~\cite{akoglu2010oddball, bindu2018discovering, niu2018visual}, or spectral analysis~\cite{chen2013novel} to detect anomalous substructures (i.e., fraud groups/events). 
However, audit experts told us that these methods are prone to false positives, due to the ambiguous nature of collusive fraud in health insurance. Excluding false positives is time-consuming for auditors and can significantly reduce detection efficiency. 
ML methods mainly use graph neural network (GNN) models to detect collusive fraud~\cite{wang2019fdgars, xu2021towards, zhong2020financial}. Fraudsters and their associations are constructed as homogeneous or heterogeneous graphs. GNNs trained on labeled data can yield the representation of fraudsters and be further applied to judge unlabeled individuals. Unfortunately, large amounts of labeled data are indispensable for high-performance GNNs. Without sufficient labeled fraud, GNN models are not applicable in our scenario.

To address these challenges, we propose a novel visual analytics approach to help health insurance audit experts identify suspicious groups, investigate the visit behavior of suspicious patients, and validate collusive fraud results.
We propose a co-visit network to represent the relationship among patients. The weights of the edges are calculated by extracting the characteristics of collusive fraudsters, such as the time gap and number of visits. Suspicious groups with multiple simultaneous visits to the same location can then be identified by a weighted community detection algorithm.
The algorithm is integrated into a prototype system, \textit{FraudAuditor}, that supports experts in interactively browsing and improving model detection results. \textit{FraudAuditor} can help experts quickly locate and examine fraud by observing co-visit links in visualizations of patient medical behavior. Combined with contextual information such as disease, drug, and fee information, false positive groups can be verified and excluded. We provide case studies and expert interviews in real health insurance scenarios to validate the effectiveness of the proposed approach.

The contributions in this work include:

\begin{itemize}[noitemsep,topsep=0pt]

  \item A problem characterization that summarizes the requirements of collusive fraud detection in the scenario of health insurance.

  \item A novel three-stage visual analytics approach to detect collusive fraud in health insurance that considers the visit pattern of fraud groups and expert knowledge.

  \item An interactive prototype system, \textit{FraudAuditor}, to facilitate the identification, examination, and validation of suspicious collusive fraud groups.
 
\end{itemize}

\section{Related Work}\label{sec:rel}


\subsection{Collusive Fraud Detection Models}

Anomaly detection models are widely applied to detect collusive fraud from graph data that records interpersonal events by identifying groups with unexpected behaviors~\cite{bindu2018discovering, molloy2016graph, li2012mining, joudaki2015using, akoglu2015graph}. Related methods can be divided into statistics-based models and ML methods.

Statistics-based models identify anomalies through the statistical information of nodes, edges, or sub-graphs. For instance, Akoglu et al.~\cite{akoglu2010oddball} extracted structural features, such as node degree or centrality, from the graph to find egonets. SpamCom~\cite{bindu2018discovering} identified spammer communities on Twitter by using structure and attribute features such as Twitter content similarity, user topology, and user profile. In healthcare scenarios, Chen et al.~\cite{chen2013novel} applied a spectrum analysis-based community detection method to detect patient referral fraud cases from a bipartite graph of physicians and specialists. Zhao et al.~\cite{zhao2019health} generated a dynamic heterogeneous information network containing patients, hospitals, and diseases. Then, they identified anomalies that fit predefined fraud patterns (e.g., the high-cost single treatment) over fixed or variable periods. Statistics-based methods can produce initial fraud candidates but may have erroneous results, requiring further validation by experts.

ML methods typically use GNN to detect fraud, as it is powerful for learning a deep representation of nodes. Previous studies are either conducted on homogeneous~\cite{wang2019fdgars, ding2019deep} or heterogeneous graphs~\cite{xu2021towards, wang2019semi, zhong2020financial}. Wang et al.~\cite{wang2019fdgars} constructed a network of reviewers in online app stores, where nodes (i.e., reviewers) are connected if they have reviewed the same app. The reviews and behavioral features of reviewers are extracted from review logs. Then, a graph convolutional network model is trained and used to detect more fraudsters based on the identified fraud. To detect collusion for fraudulent consumer loans from individuals with various roles (e.g., sellers and intermediaries), Xu et al.~\cite{xu2021towards} propose GRC, a novel GNN model, that learns representations of different types of individuals and detects loan fraud by using attention mechanisms and imposing conditional random fields. However, these ML methods are supervised or semi-supervised and thus require fraud-labeled data, which is lacking in our health insurance scenario.

Since the boundary between fraudulent and normal behavior in healthcare insurance could be unclear, automated models can hardly learn to judge correctly and achieve satisfying accuracy. Therefore, our approach integrates a graph-based detection model with a visual interface, which supports interactive data exploration, model optimization, and result validation.

\subsection{Visual Analytics Approaches for Fraud Detection}
For human-in-the-loop fraud detection, existing studies employ visual analytics to help users understand and implement detection tasks from the perspectives of individual portraits, dramatic changes, and interpersonal events.

Individual portraits include high-dimensional records, which can be described and compared by glyph representations~\cite{ko2014analyzing, cao2015targetvue, maccas2020vabank}.
TargetVue~\cite{cao2015targetvue}'s three circular glyphs depict Twitter users' communication activities, features, and social interactions. Juxtaposed glyphs allow users to compare the behaviors of different individuals and discover possible fraud, such as social bots.
To analyze and detect fraud patterns in banking transactions, Macas et al.~\cite{maccas2020vabank} offered different glyphs to characterize bank clients. Depending on the transaction amount, beneficiaries, and transaction time, the glyph has a circular or rectangular shape complemented with a series of symbols, which enhance the analyst's understanding of typical/atypical transaction profiles.

Dramatic changes are also an important point cut of fraudulent behaviors. Previous studies have designed multiple representation techniques to visualize temporal information, such as sequence visualization~\cite{DBLP:journals/ivs/MacasPM22, zhao2014fluxflow}, radial layouts~\cite{bertini2007spiralview, silva2021visualisation}, and calendar~\cite{lin2020taxthemis}, etc. FluxFlow~\cite{zhao2014fluxflow} demonstrates the impact of anomalous information (e.g., rumors) spreading through colored circles packed on a timeline. Bertini et al.~\cite{bertini2007spiralview} proposed SpiralView, which uses radar charts with spiral time axes to show how alerts change over time to detect suspicious periodic patterns. TaxThemis~\cite{lin2020taxthemis} uses calendar heatmaps to show evidence of transferring revenue through related taxpayers.

Interpersonal events can be summarized by graph visualization. For instance, 
financial fraud between buyers and sellers can be reflected by anomalous structural patterns composed of nodes and edges~\cite{didimo2011advanced}. Niu et al.~\cite{niu2018visual} used a node-link diagram to demonstrate the loan guarantee network, where each node belongs to a community defined by a random walk algorithm and is encoded with the corresponding color. In order to identify collective anomalies, Tao et al.~\cite{tao2018visual} proposed a high-order correlation graph to support analysis processes starting with an abnormal node. Corresponding nodes that contribute to the anomaly can be easily identified through the high-order correlation graph.

Our system incorporates graph and sequence visualization. To focus on collusive fraud in health insurance scenarios, our system provides richer contextual information, such as disease, drugs, and visit frequency.

\section{Domain Characterization}\label{sec:domain}

Through intensive collaboration with health insurance experts, we get access to real-world health insurance data, learn about the patterns of collusive fraud, and summarize a set of design requirements.

\subsection{Data Description}

The data used in this paper are from the local Healthcare Security Administration we collaborate with. Under experts' guidance, we excluded irrelevant fields and anonymized identity information. Two tables hold the processed data:

\begin{itemize}[noitemsep,topsep=0pt]
    \item \textbf{Patient visit table}: Rows represent patient visits. Columns include time, patient ID, medical institution, diagnosed disease, and total fee. For example, at 16:23 on August 12, 2021, patient $P_1$ went to a hospital. $P_1$ was diagnosed with hypertension and was reimbursed 36.35 yuan for drugs through health insurance. 
    
    \item \textbf{Drug table}: Doctors prescribe drugs for each patient visit. The name and dosage of each drug are recorded in this table. Note that a prescription can include multiple drugs. For example, the doctor prescribed $P_1$ antihypertensive drugs, consisting of perindopril and enalapril.
\end{itemize}

\subsection{Problem Specification}
\label{sec:pro}
Over the past year, we have worked closely with two health insurance audit experts with four years of work experience. Through multiple interviews with them, we learned about the health insurance system, examined existing fraud cases, and summarized patterns of collusive fraud behaviors.

In our scenario, a part of the medical expense can be reimbursed when insured patients pay for their drugs. Fraud occurs when insured patients sell the drugs instead of taking them after reimbursement. 
Seeking to cash out quickly, fraudsters need to gather a large number of drugs. Thus, fraudsters always collude with each other to purchase sufficient drugs within a short period. To avoid scrutiny, they prefer to visit clinics or pharmacies in poorly regulated rural or community areas. 
However, \rc{the behavior of fraudulent groups can be confused with the normal visit behavior of patient groups with chronic diseases.} For example, if the doctors treating the chronic disease see patients only at specific times, there is a high probability that certain patients will go to the same location at a similar time.
\rc{Therefore, normal patients can also have the characteristics of \textbf{spatio-temporal connections} and \textbf{group actions} similar to the collusive fraud group, i.e., patients may visit certain medical institutes together on similar cycles.}
To avoid misjudgment, \rc{it is indispensable for experts to review and validate suspicious groups by referring to the contextual information of patients' visit behavior.}

To better understand the needs of audit experts in detecting and analyzing collusive fraud, we interviewed them and summarized the current audit process into three steps.

1. Experts use the audit system (a graphical interface of the database) to learn the overall characteristics of the health insurance data, such as the cost of claims. Then, they narrow the scope of the investigation by filtering out patients of interest, such as those whose costs exceed 10,000 yuan. Since fraudsters may have multiple tricks, experts require repeated attempts to avoid omission. 
Then, experts leverage manual rules to filter patients to further identify suspicious groups. For example, experts can identify patients with an unusually high number of drug purchases over a period of time or sequentially separate out groups with a high overlap of visit locations and similar visit times \rc{(e.g., a group of five patients frequently visit a specific drugstore within 1 hour to purchase medications)}. 

2. Experts browse the list of suspicious groups to 
start with the groups with serious hazards potentially and seize the opportunity to stop loss in time. The potential hazard of a group can be estimated according to the group size and the total claim expense. Besides, experts may also merge groups with similar characteristics during browsing to improve analysis efficiency.

3. Experts judge whether a group is a collusive fraud or a false positive by examining details of patients' visit behaviors (diseases, drugs, hospitals, etc.). Highly suspicious groups will be further investigated (e.g., surveillance video inspections and phone/on-site interviews with the patients).

\subsection{Requirement Analysis}

As mentioned above, the audit process requires a large number of manual inspections, which are extremely time-consuming and labor-intensive.
Based on discussions with experts, we summarized requirements at three levels.

Users need to learn about the patients statistics and their behavior connections from the \textbf{overview} level.

\begin{compactitem}
    \item[\textbf{R1}]
    \textbf{Show attribute distribution of medical records.} An overview allows users to understand the dataset and find a starting point to detect fraud. For example, users can learn a reasonable expense range from the distribution of patient expenses. Then, the patients whose expenses exceed the threshold should be reviewed.
    
    \item[\textbf{R2}]
    \textbf{Allow flexible data filtering.} Our dataset includes many patients' visit records. Some of them do not need to be audited because of small expenses or limited numbers of visits. Filtering patients by appropriate user-specified conditions can improve analysis efficiency. 
    
    \item[\textbf{R3}]
    \textbf{Identify the behavioral connections among patients.} Connections of visiting behaviors and drug purchasing behaviors are the basis for detecting fraud groups. Thus, these connections should be identified according to expert knowledge, namely, user-specified restrictions. For instance, patients are considered to be potentially associated only if they visit the same location in less than 15 minutes more than five times.

\end{compactitem}

Further fraud detection processes should be implemented at the \textbf{group} level.

\begin{compactitem}
    \item[\textbf{R4}]
    \textbf{Detect patient groups.} Patient groups can be detected based on various user-specified rules (e.g., whether there exist specific behavior connections or whether the total expense exceeds a limit). Automation can be leveraged to guarantee the efficiency of group detection.

    \item[\textbf{R5}]
    \textbf{Support suspicious group selection.}
    Given a large number of detected candidate groups, users need to quickly locate those with the most suspicion or hazards. Recommending groups using effective ranking approaches can accelerate group selection.

\end{compactitem}

Finally, users need to check group details from the \textbf{patient} level and find evidence of fraud.

\begin{compactitem}
    \item[\textbf{R6}]
    \textbf{Support suspicious patients verification.} Understanding the intra-group similarities of patient behaviors, such as prescribed diseases, drug purchases, and selection of medical institutions, can help users exclude irrelevant patients and examine suspicious patients.
    
    \item[\textbf{R7}]
    \textbf{Visualize the visit records of an individual patient.} As mentioned earlier, auto-detection can hardly differentiate fraud groups from patients with specific visit needs, which leads to false positives. Users should examine the identified suspicious fraudsters to prove or disprove their suspicions. Visualizing patients' histories of medical visits could help users gather evidence regarding the continuity and rationality of the visits. In this way, fraudulent groups and false positive groups can be differentiated based on expert knowledge.

\end{compactitem}

\section{Our Approach}\label{sec:approach}
This section provides an overview of the visual analytics approach and introduces the two employed models.

\subsection{Approach Overview}

\begin{figure*}[ht!]
    \centering 
    \includegraphics[width=2\columnwidth]{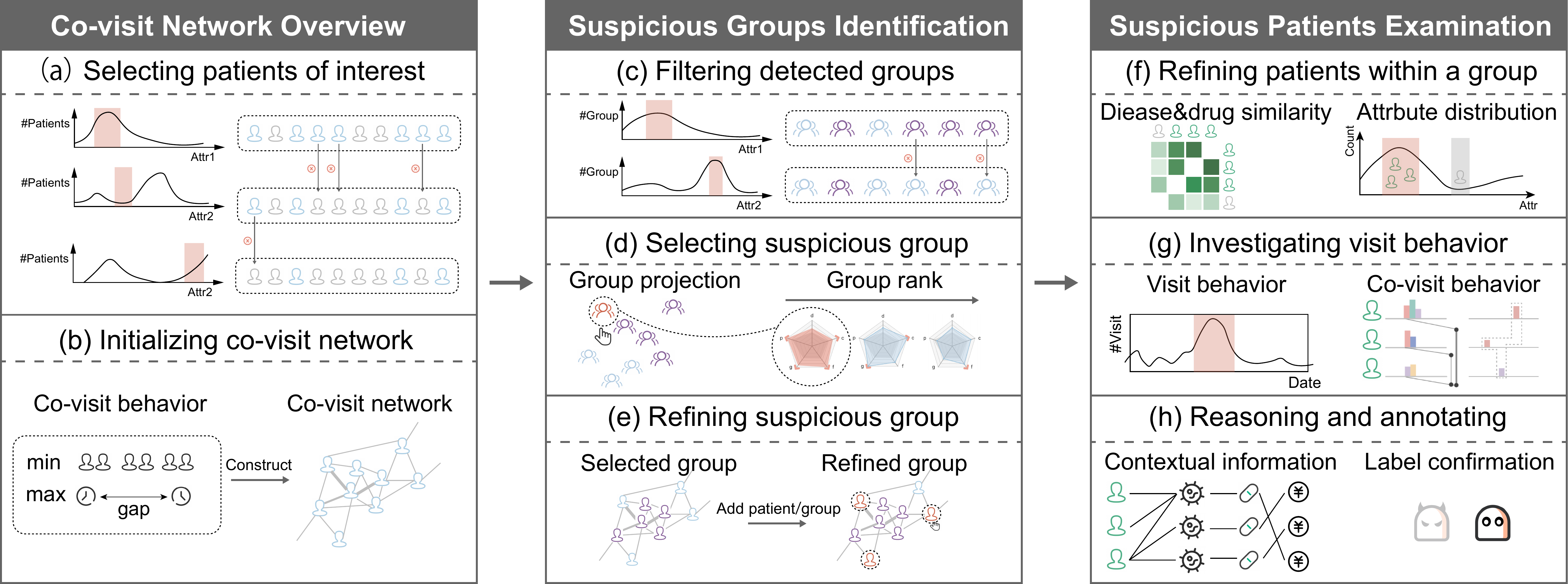}
    \caption{The three-stage approach that can help users identify, examine, and verify suspicious groups of collusive fraud in health insurance.}
    \label{fig:workflow}
\end{figure*}

Based on the design requirements elaborated in \autoref{sec:domain}, we propose a visual analytics approach (see \autoref{fig:workflow}), which enables users to analyze health insurance records at multiple levels to drill down into the data of interest, locate suspicious fraudulent groups, and find patient-level evidence to verify collusive fraud. Our approach consists of three stages: (1) co-visit network overview, (2) suspicious groups identification, and (3) suspicious patients examination.

\textbf{Co-visit network overview.} In the first stage, users seek a general understanding of the data by checking the attribute distributions (\textbf{R1}, \autoref{fig:workflow}-a). According to data distribution and domain knowledge, users then filter patients for analysis (\textbf{R2}, \autoref{fig:workflow}-a). 
Next, users check connections between the filtered patients interactively (\textbf{R3}, \autoref{fig:workflow}-b). Collusive behaviors can be disclosed by time gaps of visits or the number of co-visits. Users are supported to specify the definition of complicit behaviors by setting thresholds for time gaps and the co-visits number. 

\textbf{Suspicious groups identification.} In the second stage, our system employs a group mining method (see \autoref{sec:sgm}) to detect complicit groups according to the user-specified definition (\textbf{R4}). 
Then, our system provides multiple selection strategies to help users locate target groups from the list of detected groups (\textbf{R5}). Feasible strategies are multi-attribute filtering (\autoref{fig:workflow}-c), group comparison, and group ranking (\autoref{fig:workflow}-d). Users can also add neighboring patients or groups to optimize detection results (\autoref{fig:workflow}-e).

\textbf{Suspicious patients examination.} In the third stage, our system calculates the similarity of the prescribed diseases and drugs between each pair of patients in a group (see \autoref{sec: dds}). According to the similarity, users can assess the likelihood of collusive fraud (\textbf{R6}, \autoref{fig:workflow}-f). Patients with a low likelihood can be excluded interactively. Next, users can investigate the rest patients by inspecting their visit behaviors at different time granularities (\textbf{R7}, \autoref{fig:workflow}-g). Our system allows users to quickly understand the time periods and frequency of co-visits among them. 
We also provide contextual information, including disease, drug, and fee, to help users reason and annotate whether the suspicious behavior is collusive fraud (\textbf{R7}, \autoref{fig:workflow}-h).

\subsection{Suspicious Group Mining}
\label{sec:sgm}
\rc{To detect suspicious groups with spatio-temporal connections and group action characteristics (see \autoref{sec:pro})}, we propose a suspicious group mining method to detect collusive fraud in health insurance (\textbf{R4}). Our method first builds a co-visit network to represent the spatio-temporal relationship among patients. Based on the co-visit network, the method uses a modularity optimization-based community detection algorithm to mine suspicious groups. For clarity of description, we have listed the notations in \autoref{tab:notations}. See Algorithm~\autoref{alg:sgm} for the pseudo-code.

\begin{table}[htb]
  \centering
  \caption{Notation Definitions.}
  \label{tab:notations}
  \small
  \begin{tabular}{p{0.15\linewidth} p{0.75\linewidth}   }
  \toprule
  \textbf{Notation}  & \textbf{Description}  \\ 
  \midrule
      $\mathbf{P}$ & The patients set \\
      $m$ & The number of patients \\
      $\mathbf{V}$ & The visits set \\
      $n$ & The number of visits \\
      $t_i$ & The time of visit $v_i$ \\
      $\theta_1$ & The maximum time gap for a co-visit \\
      $\theta_2$ & The minimum number of co-visits \\
      $\mathbf{CV}(p_i,p_j)$ & The co-visit behaviors between patient $p_i$ and $p_j$\\
      $w(v_i,v_j)$ & The weight of a co-visit about visit $v_i$ and $v_j$ \\
      $w(p_i, p_j)$ & The weight of co-visits between patient $p_i$ and $p_j$ \\
      $\mathbf{W}$ & The weight between patients \\
      $\mathbf{D}$ & The diseases set \\
      $\mathbf{C}$ & The number of visits for each disease in $\mathbf{D}$ \\
      $w(d_i)$ &  The contribution of disease $d_i$ to the similarity of patients $p_i$ and $p_j$ \\
      $c_i$ &  The number of visits of disease $d_i$ \\
      $sim(p_i, p_j)$ &  The similarity of patient $p_i$ and $p_j$ \\
  \bottomrule
  \end{tabular}
\end{table}

\textbf{Co-visit network construction}. Patients in a collusive fraud group frequently visit the same medical institution within relatively short time periods. Considering such a characteristic, we construct a co-visit network $\mathbf{G}$ among patients to summarize the co-visit behaviors and detect collusive fraud. A node in the network represents a patient. An edge between two patients records the co-visit behaviors between the two patients. \rc{If the medical institutions of the two corresponding visits of two patients are the same, and the time gap is less than a threshold $\theta_1$ (the default is 1 hour, which can be adjusted to 6, 12, or 24 hours), it is considered a co-visit.} For patients $p_i$ and $p_j$, their co-visit behaviors are represented as
$\mathbf{CV}(p_i, p_j) = \{(v_{i1}, v_{j1}), \cdots, (v_{is}, v_{js})\}$ and $s$ is the total number of visits they made together. 

\textbf{Edge weight calculation}. The edge weight indicates the likelihood that the two patients belong to the same group. We calculated the weights of the edge $w(p_i, p_j)$ based on the number of co-visits and the visiting time gap. 
As shown in the \autoref{eq:1}, the weight of a co-visit is inversely proportional to the visit time gap. To avoid the impact of occasional visits with a small time gap on the weight, \rc{inspired by the ReLU activation function, we set the cutoff time to 10 minutes based on expert experience, and weights less than that interval are considered to be the same.}


\rc{
\begin{equation} \label{eq:1}
    w(v_i, v_j)=\left\{
    \begin{array}{rcl}
        \frac{1}{max(10 \, \rm minutes, |t_i - t_j|)} &  & {|t_i - t_j| \leq \theta_1} \\
        0                                             &  & {otherwise}
    \end{array} \right.
\end{equation}
}

The edge weight $w(p_i, p_j)$ between two patients is the total of their co-visit weights, defined as \autoref{eq:2}.
An adjustable threshold $\theta_2$ (default as 4) for the minimum number of co-visits is set here to avoid random factors. The co-visit weight being less than the threshold indicates the low probability of both belonging to the same group.

\begin{equation} \label{eq:2}
    w(p_i, p_j)=\left\{
    \begin{array}{rcl}
        \sum_{z=1}^{|\mathbf{CV}(p_i, p_j)|} w(v_{iz}, v_{jz}) &  & {|\mathbf{CV}(p_i, p_j)| \geq \theta_2} \\
        0                                             &  & {otherwise}
    \end{array} \right.
\end{equation}

\textbf{Community detection}. \rc{In order to mine suspicious groups from the co-visit network, we use Louvain~\cite{blondel2008fast}, an community detection algorithm based on modularity optimization. The algorithm is applicable to weighted graphs and supports the exclusion of non-community nodes, which can yield clear detection results since most patients in the healthcare scenario are normal.}

\begin{algorithm}[]
    \caption{Suspicious Group Mining}
    \small
    \label{alg:sgm}
    {{
                \begin{algorithmic}[1]
                    \Require
                    $\mathbf{P}$: the patient set;
                    $\mathbf{V}$: the visit records;
                    $\mathbf{W}$: the weight between patients;
                    $\theta_1$: the maximum time gap;
                    $\theta_2$: the minimum co-visit times.
                    \Ensure
                    $\mathbf{G}$: the co-visit network;
                    $\mathbf{SG}$: the suspicious groups.
                    \State $\mathbf{CV} \gets$ extract co-visit behavior from $\mathbf{V}$
                    \State $\mathbf{W} \gets \mathbf{0}$
                    \For{each patient pair $(p_i, p_j)$}
                    \State $w(p_i, p_j) \gets 0$
                    \If {$|\mathbf{CV}(p_i, p_j)| \geq \theta_2$}
                    \For{each co-visit ($v_{ik}, v_{jk}$) in $\mathbf{CV}(p_i, p_j)$}
                    \State $w(v_{ik}, v_{jk}) = \mathbbm{1}(|t_i - t_j| \le \theta_1)\frac{1}{max(10 \, \rm minutes, |t_i - t_j|)}$
                    \State $w(p_i, p_j) \mathrel{+}= w(v_{ik}, v_{jk})$
                    \EndFor
                    \EndIf
                    \EndFor
                    \State $\mathbf{G} \gets (\mathbf{P}, \mathbf{CV}, \mathbf{W})$
                    \State $\mathbf{SG} \gets \Call{cdlib.algorithms.louvain}{\mathbf{G}, \mathbf{W}}$ 
                    \State \Return {$\mathbf{G}, \mathbf{SG}$}
                \end{algorithmic}}}
\end{algorithm}

\subsection{Disease and Drug Similarity} \label{sec: dds}

To verify collusive fraud groups, we need to calculate the similarity among patients based on their prescribed diseases and corresponding drugs (\textbf{R6}). At first, we tried to calculate the similarity according to the string texts of diseases and drugs, but the results were not satisfactory. For example, such a calculation would lead to the headache being similar to the stomachache and not similar to the stroke, when in fact both headaches and strokes are brain disorders with a closer relationship. Later, we found that either diseases or drugs have hierarchical encoding, which can reflect the similarity information. The ICD10~\footnote{\url{https://en.wikipedia.org/wiki/ICD-10}} coding of diseases and the standard coding of drugs~\footnote{\url{https://code.nhsa.gov.cn/toDetail.html?infoId=5546&CatalogId=2}} encode diseases and drugs hierarchically by large, medium, and small class.
For example, diseases J11 (influenza) and J18 (pneumonia) are similar, but they are very different from M54 (back pain).

Hence, we propose a nearest-match-based similarity calculation method that considers the disease/drug codes and the number of visits for the corresponding diseases. For each disease/drug, we need to find the most similar one in another \rc{patient's disease/drug set, so it is not affected by the specific order of visits}. Assume that a patient has been prescribed several diseases $\mathbf{D} = \{d_1, d_2, \cdots, d_l \}$. The corresponding numbers of visits for each disease are $\mathbf{C} = \{c_1, c_2, \cdots, c_l\}$. 

\begin{figure}[h]
    \centering 
    \includegraphics[width=\columnwidth]{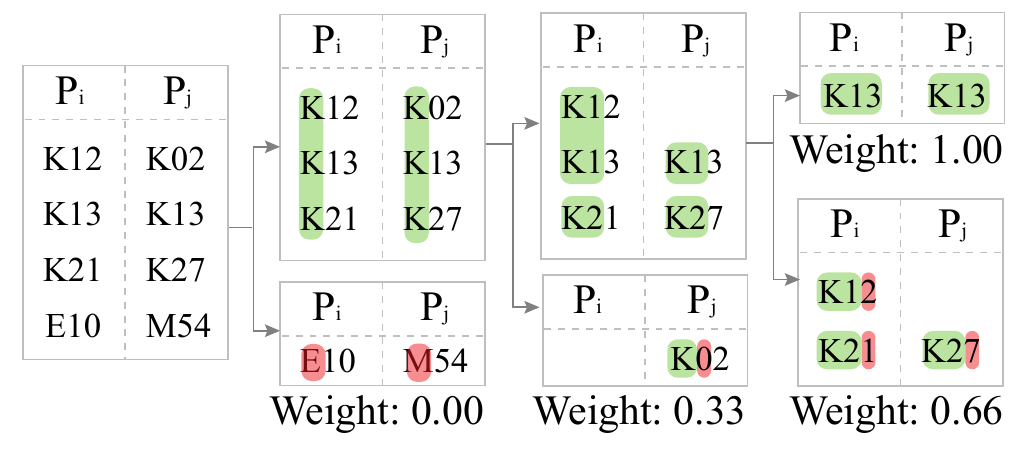}
    \caption{An example of disease weight calculation.}
    \label{fig:algorithm}
\end{figure}

As is shown in \autoref{fig:algorithm}, from the first letter of $p_i$'s and $p_j$'s diseases, we can see that both of them have been treated for diseases beginning with ``K''. However, $p_i$'s E10 (i.e., Type 1 diabetes), and $p_j$'s M54 (i.e., back pain) are not shared with each other. Thus, E10 and M54 contribute no disease similarity. Next, according to the code of the second letter, the disease set beginning with K can be divided into three medium-class sets: K0, K1, and K2. The K02 disease of $p_j$ does not correspond to the remained diseases of $p_i$, while the remaining diseases are further calculated according to the letters of the third letter until the diseases of the two patients in the set are exactly the same. The closer the two diseases are, the greater their contribution to the calculation of similarity. Therefore, the weight of each disease is determined by the longest prefix of the set in which it last stays, having

\begin{equation}
    w(d_i)= \frac{\text{the length of the prefix letter}}{\text{the total length of the coded letter}}.
\end{equation}

For example, the weight of disease K02 is 1/3$\approx$0.33, K12 is 2/3$\approx$0.66, and K13 is 1.00. The same goes for drugs. The number of medical visits for diseases/drugs also reflects their similarity. The similarity between two patients is calculated as follows:

\rc{
\begin{equation}
    sim(p_i, p_j) = \dfrac{\sum w(d_k^{p_i}) \cdot c_k^{p_i} + \sum w(d_k^{p_j}) \cdot c_k^{p_j}}{\sum c_k^{p_i} + \sum c_k^{p_j}}.
\end{equation}
}

\section{System Design}\label{sec:design}

To help users implement the approach mentioned in \autoref{sec:approach}, we developed an interactive prototyping system, \textit{FraudAuditor}. This section presents a system overview and introduces the details of the visual design and interactions.

\subsection{System Overview}

\begin{figure*}[ht!]
    \centering 
    \includegraphics[width=\linewidth]{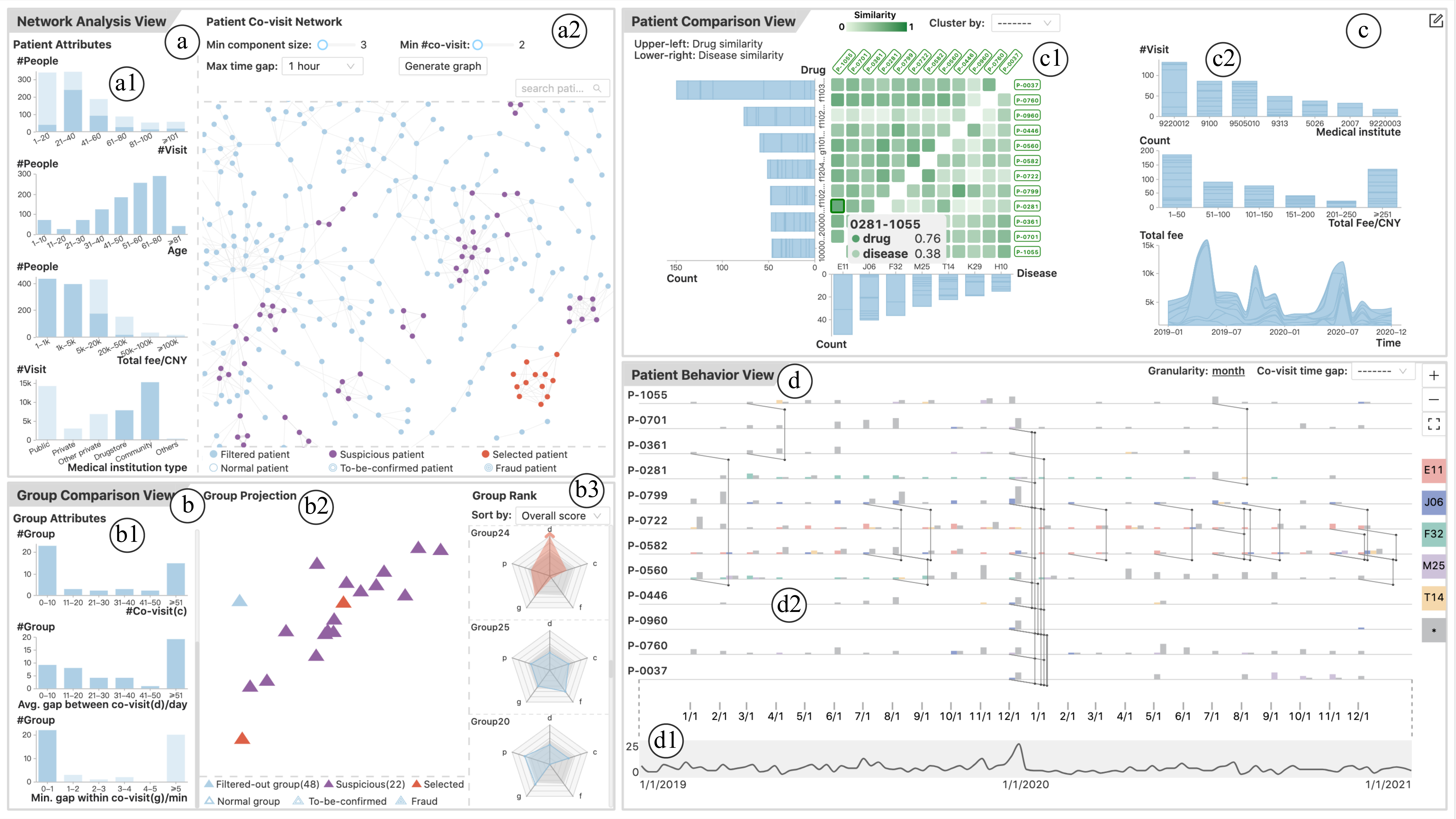}
    \caption{\textit{FraudAuditor} facilitates the identification, examination, and annotation of collusive fraud in health insurance. (a) The network analysis view supports interactive filtering of patient attributes and constructing a co-visit network. (b) The group comparison view provides interactive filtering, similarity analysis, and ranking of groups. (c) The patient comparison view helps users analyze the similarity and distribution of diseases, drugs, and other attributes among patients within a group and helps select suspicious patients to be analyzed. (d) The patient behavior view supports the inspection and annotation of detailed patient visit records and the analysis of co-visit behaviors.}
    \label{fig:teaser}
\end{figure*}

The system contains four views, as shown in \autoref{fig:teaser}: the network analysis view, the group comparison view, the patient comparison view, and the patient behavior view. We describe an analysis flow to demonstrate how these four views help the user discover, analyze, and validate suspicious groups of collusive fraud based on health insurance data. The user first learns the data distribution from the bar chart of patient attributes in the network analysis view (\textbf{R1}), based on which he can interactively filter the data of interest (\textbf{R2}). Then he sets parameters on co-visit behavior and generates the co-visit network between patients in the patient co-visit network view (\textbf{R3}). The results of the automatic detection model are also displayed in the network in real-time by highlighting. The attribute distribution, similarity, and ranking of detected groups can be viewed in the group comparison view (\textbf{R5}). He clicks on the top-ranked group, and its position in the network is highlighted simultaneously. In the patient comparison view, he compares different patients in the group using the disease and drug similarity matrix, stacked bar chart, and area chart (\textbf{R6}). From there, he selects several suspicious patients and goes to the patient behavior view for further investigation. He visually analyzes the visit pattern and co-visit distribution of patients from the visit sequence visualization. Combined with the visualization of contextual information such as diseases and drugs, he infers and labels these patients as engaging in collusive fraud (\textbf{R7}).

\subsection{Network Analysis View}

The network analysis view (\autoref{fig:teaser}-a) has two parts: (1) The patient attributes view gives an overview of patients by showing attribute distributions and supports interactive filtering of data of interest (\textbf{R1, R2}). (2) The patient co-visit network lets users interactively define co-visit behaviors, browse the resulting co-visit network, and visually inspect suspicious groups detected by automated algorithms (\textbf{R3}).

In the patient attributes view (\autoref{fig:teaser}-a1), bars indicate the distribution of patient attributes, including the distribution of patients in terms of the number of visits, age, and total fee, as well as the number of visits to different medical institutions. At first, all patients are selected, and users can click on a bar to deselect/re-select the corresponding patients. Patients that are not selected are represented by a translucent background, and a mouse hovering over the corresponding bar will bring up a tooltip showing the total number of patients belonging to the original and current patients in the interval, making it easy for users to compare the distribution of patients under different filtering conditions.

In the patient co-visit network (\autoref{fig:teaser}-a2), the control panel above allows users to interactively configure the definition of co-visit behavior. A slider controls the minimum number of co-visits, and a drop-down menu sets the maximum time gap, such as 1 hour, 6 hours, etc.
Users can browse the co-visit network by clicking "Generate Graph" and iteratively changing the co-visit definition or filtering conditions in the patient attributes view if the results are unsatisfactory.
Suspicious groups detected by the automated algorithm are displayed in the network simultaneously and marked in purple, where small groups can be filtered by adjusting the minimum component size slider. \rc{The node-link diagram was applied to visualize the graph. Because node-link diagram performs well in showing direct and indirect relationships among patients, which can support tasks on connection analysis and topology exploration.} Each node in the network represents a patient, and the edge between two nodes reflects their co-visit relationship, whose width represents the strength of the co-visit relationship (see \autoref{sec:sgm}). The network supports zooming and panning for navigation. Patients are highlighted, and patient ID and group ID (if any) are displayed when the mouse hovers over a node. If the node belongs to a group, all nodes in that group are highlighted. 
To help users understand the analysis provenance, selected, suspicious, normal, and other nodes are mapped to different visual encodings.

\subsection{Group Comparison View}

The group comparison view (\autoref{fig:teaser}-b) consists of three parts: (1) The group attributes view provides an overview of group-level attributes as well as interactive filtering capabilities. (2) The group projection supports similarity analysis of groups (3) The group rank allows sorting groups across multiple dimensions. Through initial filtering and careful selection, users can identify suspicious groups that need to be focused on for analysis (\textbf{R5}).

The group attributes view (\autoref{fig:teaser}-b1) uses bar charts to show the distribution of groups on various metrics, including the number of patients (p), the total fee per capita (f), the number of co-visits (c), the average number of days between multiple co-visits (d), and the minimum time gap within a co-visit (g), which are common metrics used by health insurance audit experts to evaluate how suspicious a group is. The view also supports interactive filtering to help narrow down the groups to be analyzed.

The group projection (\autoref{fig:teaser}-b2) maps groups to a two-dimensional plane, helping users to compare similarities and differences between groups and to detect clusters or outliers that are worthy of analysis. Using the group-level features mentioned above, the original groups are represented as a set of feature vectors. \rc{To compare the distribution of groups over different features, we used kernel PCA~\cite{scholkopf1998nonlinear} because it maintains the covariance of the data and is able to handle linearly indistinguishable cases.}
In the projection result, each triangle represents a group, and the distance between them reflects their similarity to some extent. Depending on the group status, e.g., filtered-out or user-selected, different visual encoding is applied to the group. The group that needs further investigation can be selected by clicking on it.

\begin{figure}[tb]
    \centering 
    \includegraphics[width=\columnwidth]{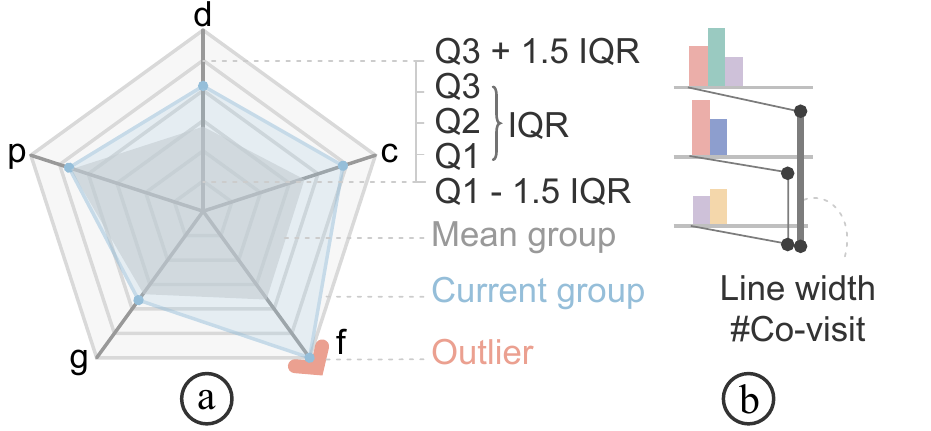}
    \caption{Explanation of visual design. (a) Customized radar chart. Five axes represent different group metrics. The darker gray area represents the average of all groups on these metrics, and the blue area represents the current group. If the value on one axis exceeds $Q3 + 1.5\mathrm{IQR}$, it is marked with an arrow as an outlier. (b)The co-visit link of note metaphor, and the width represents the number of co-visits.}
    \label{fig:radar}
\end{figure}

The group rank view (\autoref{fig:teaser}-b3) provides a drop-down menu for selecting the ranking keyword, where single group metrics or overall scores can be used.
We created a customized radar chart to visually compare groups across multiple criteria.
In \autoref{fig:radar}-a, each axis represents the metric mentioned above. Five gray lines run through each axis from inside to outside, statistically representing the lower fence, first, second, third quartile, and upper fence~\footnote{\url{https://en.wikipedia.org/wiki/Quartile}} of each group on each metric. For comparison, data transformations (e.g., adding negative signs) make all metrics more anomalous as they grow. A gray area indicates each metric's average value for reference.
The blue area shows the current group, whose size visually reflects the group's degree of suspicion.
A special arrow alerts users to outliers above the upper fence.
Hovering over this glyph displays details of the group and each metric. 
Users can click to select the corresponding group for subsequent analysis.

\subsection{Patient Comparison View}

The patient comparison view (\autoref{fig:teaser}-c) has two types of visualizations: (1) A similarity matrix of diseases and drugs that supports comparisons between patients and helps users determine whether diseases and drugs correspond to each other. (2) Stacked bar charts and area charts of the patient attribute to help analyze the contribution of different patients to the group. Through similarity and attribute comparison, users can identify suspicious patients (\textbf{R6}).

\rc{To further select suspicious patients and exclude innocent ones, users need to study the similarity of disease and drugs within a group. We show the similarity between each pair of patients in a matrix} (\autoref{fig:teaser}-c1). Cells are divided into two categories, with the upper left corner representing drug similarity and the lower right corner representing disease similarity. Both the horizontal and vertical axes of the matrix are the patients within the selected group and have the same order. Each cell is color-coded in green with the degree of similarity between the corresponding patients; the darker, the more similar. When the mouse hovers over a cell, the specific values of diseases and drug similarity and the corresponding patient IDs are displayed, and the diagonally symmetrical cell is also highlighted for comparison. To discover patterns of patient clustering on diseases and drugs, the system supports matrix reordering, where hierarchical clustering methods are used to determine new patient orders. The patient can be selected for a more detailed analysis by clicking on labels. \rc{Since there are usually at most a dozen fraudsters in a group, the matrix is less prone to visual clutter.}

\rc{If suspicious patients are found (e.g., with similar diseases but large differences in drugs), users need contextual information to further assess the rationality of drug purchase behavior.} In the stacked bar chart and stacked area chart (\autoref{fig:teaser}-c2), users can study contextual information about diseases, drugs, medical institutes visited, total fees, etc., in order to verify suspicious patients. The data corresponding to patients selected by users is highlighted, while the data for unselected patients becomes translucent to provide a clear picture of the proportion of these patients in the overall group on different attributes. When the mouse hovers over a bar or area, information about each selected patient and the entire group in that attribute interval is displayed in a tooltip.

\subsection{Patient Behavior View}

The patient behavior view (\autoref{fig:teaser}-d) contains: (1) a line chart of the number of visits over time for locating anomalies and navigation; and (2) a visualization of the patients' visit sequence that shows the evolution of patients' visits and highlights co-visits. Users can analyze visit behavior at the different temporal granularity and combine rich contextual information and domain knowledge to reason and verify whether it is collusive fraud (\textbf{R7}).

The line chart (\autoref{fig:teaser}-d1) at the bottom reflects the temporal change in the number of visits to support users in locating time periods with significant fluctuations and detecting periodic visits, etc. The gray boxes represent selected time periods and allow for range swiping and panning.

In the visit sequence visualization (\autoref{fig:teaser}-d2), each timeline corresponds to one patient's visit history during the selected time period. Different time periods lead to different view granularities, such as a month, week, day, etc., which helps improve readability and reduce the cognitive load. The bars on the timeline represent the patient's visit behavior, where the position encodes the visit time, the color represents the disease type, and the height refers to the number of visits for the corresponding disease. This allows users to quickly grasp information about the patient's main diseases, frequency of visits, etc. Since there are many possible disease types, to avoid visual clutter, we give different colors to the top 5 diseases, while the rest are represented in gray. Users can view specific information about diseases through the legend on the right side. The number of patients in the visible area can be adjusted using the plus or minus buttons on the right, and clicking the full-screen button displays all selected patients in the current window.

To represent the co-visit behavior between patients, we designed a co-visit link to explicitly show this suspicious behavior. As shown in \autoref{fig:radar}-b, if there is co-visit behavior between two patients, we extend a line at the corresponding position of each of the two patients' timelines and link them to each other with a vertical line. In addition, since the current timeline may contain aggregated visit events, the width of the vertical line is used to encode the number of co-visits. We did not use arcs to connect co-visit behaviors because they tend to cause more crossings and visual clutter.

To support an in-depth analysis of co-visit behavior, users can select the threshold in the co-visit time gap drop-down box to switch to the co-visit view (see \autoref{fig:case1}-e) and examine all co-visit behaviors within the selected time period.
When the mouse hovers above the bar, further contextual information about this co-visit appears, including each patient's visit time, medical institution, list of drugs, etc. Users can check a patient's age, gender, disease, and drug costs by clicking on the patient ID.
This supporting information helps users reason and verify if they are committing collusive fraud. If the review is complete, users can click the pencil button in the upper right corner to annotate these patients with a reason on the pop-up labeling page.

\begin{figure}[tb]
    \centering 
    \includegraphics[width=0.9\columnwidth]{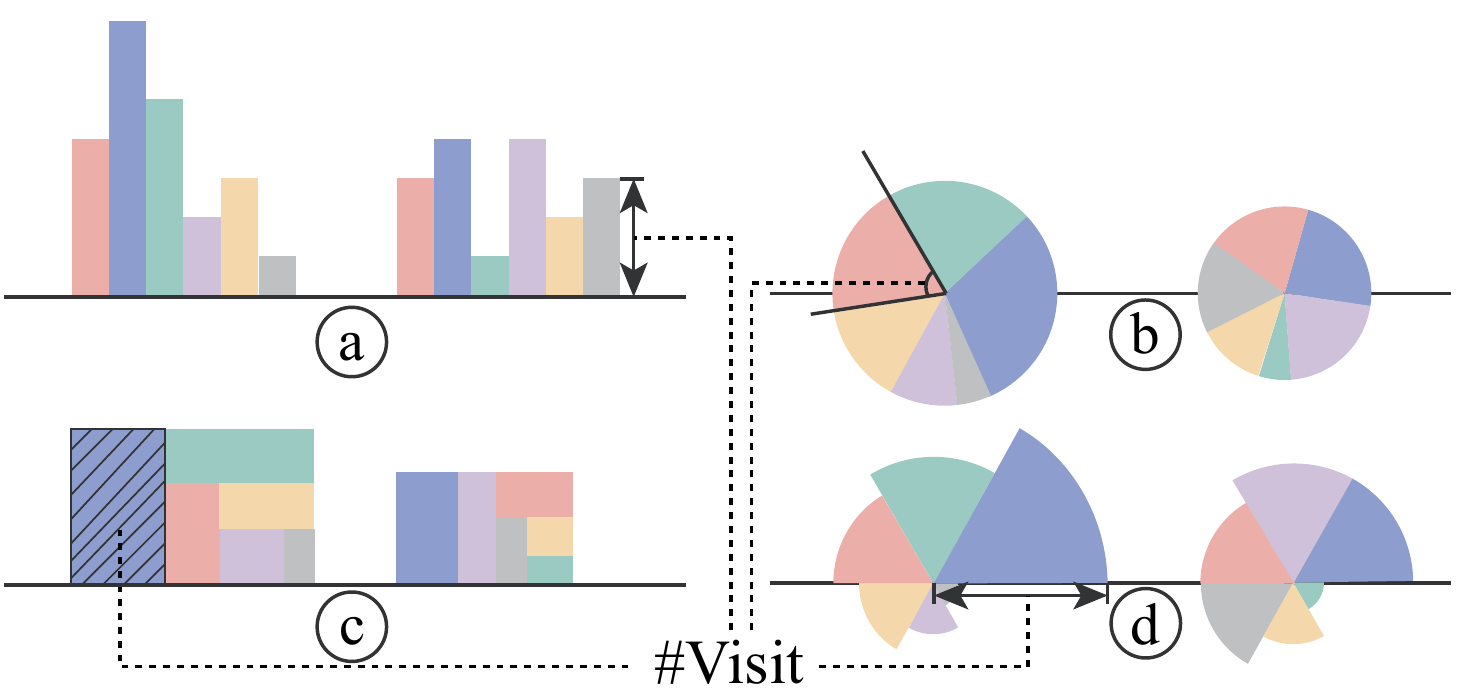}
    \caption{Alternative design of patient medical visit behavior. The number of visits is indicated by (a) the height of the bar, (b) the angle of the sector, (c) the area of the rectangle, and (d) the radius of the sector.}
    \label{fig:alter}
\end{figure}

\textbf{Alternative design.} Before we finalized the bar chart for the patient visit record, we had three alternative designs: the pie chart, the treemap, and Nightingale's rose chart. In \autoref{fig:alter}, the number of visits is denoted by height, angle, area, and radius, respectively. Although the latter three representations are more compact, the visual elements representing each disease have different angles, making comparisons difficult. Also, for the pie chart and Nightingale's rose chart, the radius and area are squared, which can easily mislead users. The bar chart, on the other hand, has a fixed orientation, which is suitable for disease comparison.


\section{Evaluation}\label{sec:eval}

To demonstrate the effectiveness of \textit{FraudAuditor}, we conducted two case studies and expert interviews using the real-world health insurance dataset. We sampled the records of 1,035 patients in a district from 2019 to 2020, which contained more than 46,000 visit records and more than 300,000 drug records. \rc{In actual scenarios, experts will select data of similar scale for analysis by spatio-temporal filtering.}

\subsection{Case Studies}

\subsubsection{Examining Crafty Fraud Group}

\begin{figure*}[htb!]
    \centering 
    \includegraphics[width=2\columnwidth]{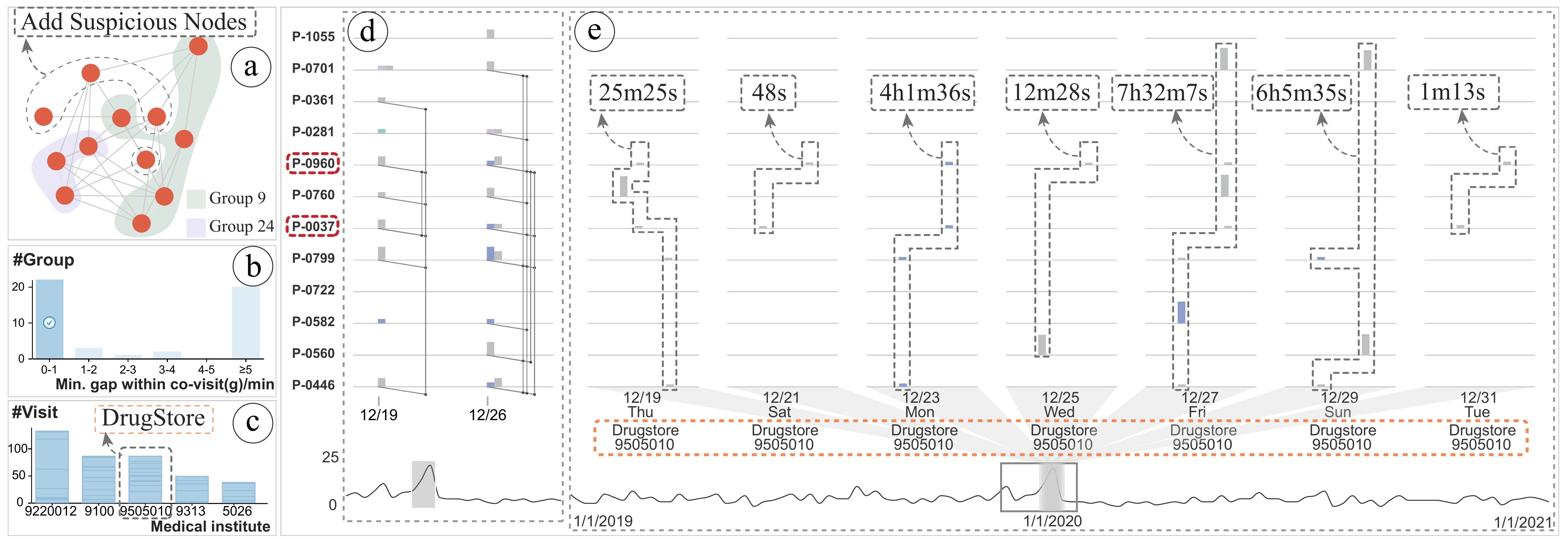}
    \caption{The visual analysis process of case 1 mainly consists of (a) selecting two suspicious groups and their neighbors, (b) filtering groups, (c) browsing patients' attributes, (d) checking frequent co-visit records of short duration, and (e) verifying co-visit behaviors in multiple time periods.}
    \label{fig:case1}
\end{figure*}

A health insurance auditor wanted to detect complex patterns of fraudulent behavior. As shown in the \autoref{fig:teaser}-a1, he chose the drugstore and community hospital in the patient attributes view, as they are poorly regulated and fraud-prone.
He noticed a significant decrease in the number of patients with 1-20 visits. The majority visited drugstores and community hospitals concentrated in the range of 21-60 times. 
Based on the prior knowledge, he set the maximum time gap for a co-visit to 1 hour and the number of co-visit to at least 2.
To avoid missing collusive fraud, while excluding single patients and small groups of only two patients, he set the minimum group size as 3 and adjusted the minimum number of co-visits from 4 to 2 to generate the co-visit network. He then clicked on the button to generate the co-visit network and browsed the results of the collusive fraud detection models, which contained 48 groups.

As shown in \autoref{fig:case1}-b, to filter the groups with short visit intervals, he selected the 0-1 minute bar in the chart about the minimum gap within the co-visit in the group attributes view.
In the remaining groups, he noticed that the two groups in the patient co-visit network are very closely connected (\autoref{fig:case1}-a). Four surrounding points connected to them aroused the expert's interest and were therefore added for analysis.
In the patient comparison view, he noticed that the selected patients had low disease similarity and high drug similarity (\autoref{fig:teaser}-c). To understand why diseases and drugs did not correspond, he selected all the patients and found that their fees fluctuated wildly.

He then went to the patient behavior view to investigate the details of visits. He noticed a spike in the timeline of the number of visits and dense co-visit links in the visit sequence visualization around December 2019 (\autoref{fig:teaser}-d).
Patients P-0037 and P-0960, in particular, had no medical records from January 2019 to November 2019 and rarely visited medical institutions after 2020. 
Their first medical visit was a co-visit in December 2019.
So he narrowed down the selected time range (\autoref{fig:case1}-d).
After adjusting the time granularity to a week, he noticed multiple co-visits in the weeks of December 19 and 26, indicating suspicious frequent co-visits in a short period.
Such frequent co-visits in a short time were very suspicious.
So he adjusted the co-visit time gap to 15 minutes and found that the co-visits of these patients all took place in Institution 9505010 (\autoref{fig:case1}-e), which was the drugstore they often visited (\autoref{fig:case1}-c).
He hovered over the bar of co-visit to check their specific time of visits and found that their time for medicine purchase was very close, and many times within 1 minute. 
The cost of each visit was also within a certain range.
He clicked on patients to view the specific information about diseases and fees.
Results showed that all the diseases had no specific directivity and were messy, and the cost of each visit was relatively fixed. 
Therefore, he speculated that these patients were likely a collusive fraud group and tagged with the corresponding labels.
The subsequent investigation confirmed that these fraudsters collude with salespeople to make money by frequently buying drugs to resell while also helping the drugstore increase sales for kickbacks.


\begin{figure}[tb]
    \centering 
    \includegraphics[width=\columnwidth]{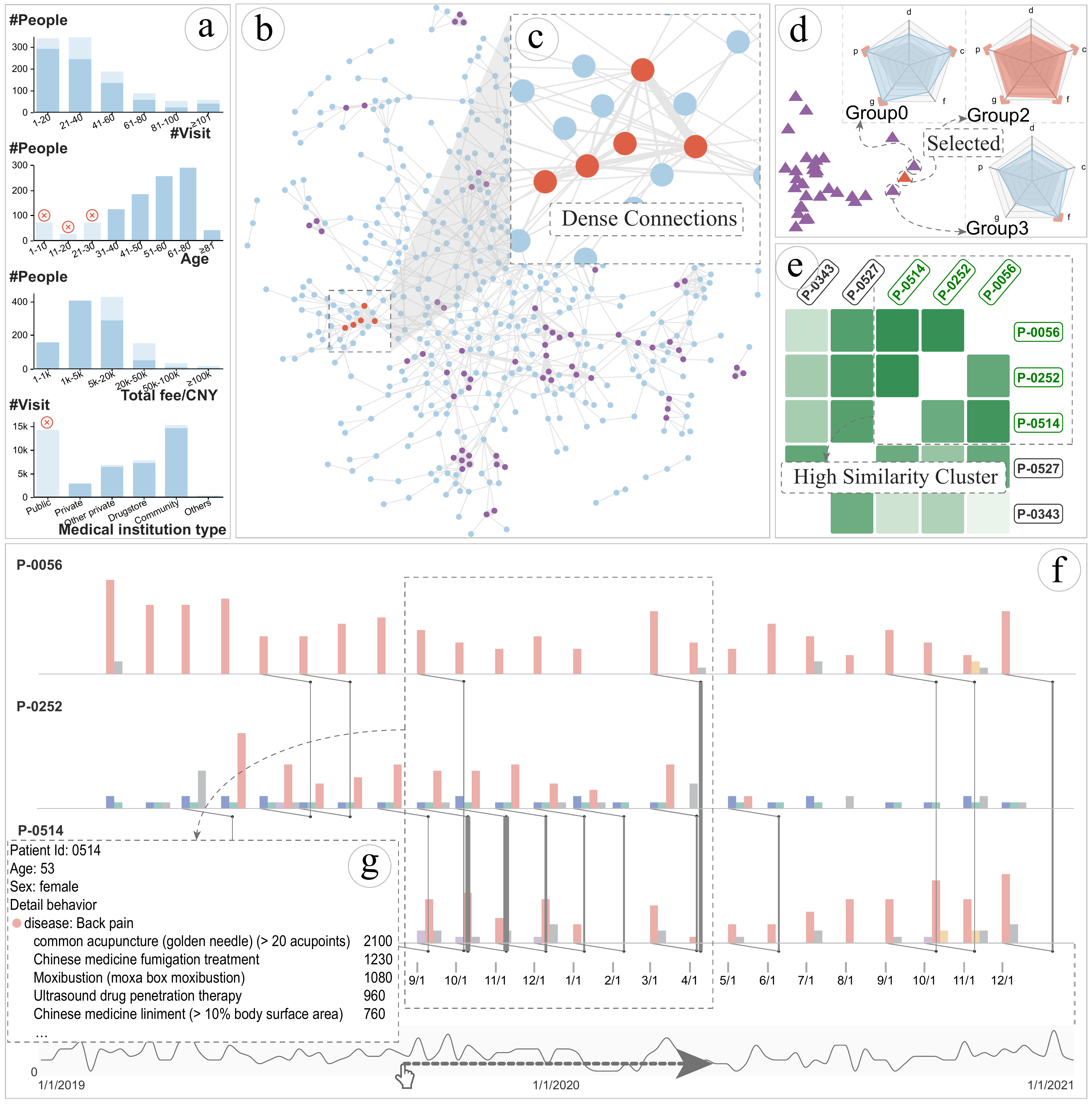}
    \caption{The visual analysis process in case 2 mainly consists of (a) filtering patients based-on attributes, (b) detecting groups, (c) identifying groups based on connections, (d) selecting suspicious groups, (e) comparing intra-group similarity, (f) exploring detailed patient behaviors, and (g) reasoning through contextual information.}
    \label{fig:case2}
\end{figure}

\subsubsection{Excluding False-positive Group with Chronic Diseases}

Another health insurance auditor wanted to exclude false-positive groups, misclassified as collusive fraud because of collective visit behavior for chronic diseases.
After reviewing the patient attributes view, she unchecked patients younger than 30 years old who were less prone to chronic disease.
The bar chart on the number of visits and the number of people showed that the majority of patients with more than 60 visits were the patients retained after filtering and were also the primary visitors to various medical institutes.
After understanding the basic characteristics, she proceeded to observe the co-visit network between these patients.
The resulting network showed that almost everyone was connected.
She attributed this mainly to the fact that the data contained numerous public hospital visits, making it easy for patients to have co-visits with others.
\rc{She, therefore, excluded records of visits to public hospitals (visits to other institutions were still retained) (\autoref{fig:case2}-a) because the large number of visits to these hospitals could easily lead to connections between patients and the likelihood of fraud is lower in well-regulated institutions.} The regenerated network had significantly fewer edges (\autoref{fig:case2}-b), and suspicious groups marked in purple were clearly visible. 

She then found some outliers in the group projection results.
She hovered over one outlier and found that it ranked \#2 on the group list, and had 4 anomalous metrics in the radar chart (\autoref{fig:case2}-d), such as 140 co-visits.
So she clicked on that radar chart and decided to further analyze the patients within that group.
Using the disease bar chart, she found that the primary diseases in this group were chronic diseases such as back pain and hypertension.
The top five drugs purchased included acupuncture, traditional Chinese fumigation medicine, and Qianghuo~\footnote{\url{https://en.wikipedia.org/wiki/Notopterygium_incisum}}, consistent with chronic diseases.
In the matrix view, after clustering the patients by disease, she found that P0056, P0252, and P0514 had a high disease and drug similarity (\autoref{fig:case2}-e), with P0514 and P0056 having up to 90\% disease similarity and the rest having more than 70\% disease similarity between them.
She then selected these three patients, and the highlighted area in the bar chart showed their distribution in each attribute relative to the entire group. Most of these patients' visits were distributed in a community hospital numbered 9202.
The area chart about fees reflected that their medical insurance expenses are relatively stable.

She suspected the three patients might have been mistakenly detected as a collusive fraud group, so she moved to the patient behavior view to investigate visit behavior for further confirmation.
She found that the pink bars representing back pain dominated the visit sequence visualization and appeared at regular intervals (\autoref{fig:case2}-f).
Between September 2019 and March 2020, there were multiple co-visit links between P0252 and P0514, so she swiped the timeline to this period.
From the co-visit view, she learned that these patients visited the community hospital (ID 9202) every week or so. The height of the bar also showed that they spent about the same amount each time, consistent with the usual regular drug change cycle.
Clicking on patients P0252 and P0514 revealed that their ages and total fees were similar (\autoref{fig:case2}-g), so she inferred that they might be patients who knew each other and went to the doctor together.
So she marked the group as normal in the annotation view and gave the corresponding reason.

\subsection{Expert Interview}

To assess the usefulness and effectiveness of \textit{FraudAuditor}, we conducted interviews with six domain experts. These experts are our collaborators but have not been involved in the design process. Two of them (E1, E2) are experts in health-related products and are familiar with the actual scenarios and business of health insurance.
The other four (E3-E6) are experts in health insurance fraud detection, understanding common fraud patterns, and detection algorithms. 

\textbf{Procedure.} Each interview lasted for 90 minutes. \rc{First, we started with a 30-minute training on \textit{FraudAuditor}'s purpose and usage flow.} Then the experts spent 45 minutes exploring and discovering suspected group fraud using the system. In the last 15 minutes, we asked each expert to complete a questionnaire consisting of quantitative evaluations based on a five-point Likert scale and qualitative questions. We summarize the insights in the following four aspects.

\textbf{Suspicious group mining model.} 
Experts agree that the detection algorithm, considering both the number of co-visits and the time gap of a co-visit, can identify suspicious groups more accurately (ratings: 4/5). Experts said most existing automated algorithms were black boxes without mechanical explanation. With \textit{FraudAuditor}, experts can understand the detection mechanics and interactively adjust them (e.g., setting the number of co-visits and the time gap threshold of a co-visit based on their own experience).
The effects of different algorithm parameters on the results are visible in real-time, which can boost experts' trust in the algorithm.
E3 said, \textit{``The co-visit network can help verify many of my hypotheses. For example, public hospitals cause many co-visits, and excluding records in them wouldn't have much effect on the numbers of groups.''} E3 added that \textit{``if the algorithm can take into account the continuity of visits, it will be more effective.''}

\textbf{Visualization and interaction.} Experts believe \textit{FraudAuditor} can meet the needs of analysis. 
\textit{FraudAuditor} uses basic visualizations (e.g., bar charts, line charts, and node-link diagrams) and common statistical metrics (e.g., median and mean).
Since healthcare experts are not familiar with visualization, these simple visual designs lower the learning barrier.
E1 mentioned that \textit{``the patient behavior view (ratings: 4.5/5) is beautiful and practical in design, which can help assess the evolution of the disease and co-visit behavior.''}
E4 noted that \textit{``it takes some exploration to understand the coordination of the system. I never used similar functions before and gradually got familiar with them after a period of use.''}
E2 stated that \textit{``usually we don't have much time to verify each group in detail in real work.
The interactive filtering of patients (ratings: 4.3/5) and groups in the system, as well as the ranking in the customized radar chart (ratings: 4/5), enabled me to select some groups of higher suspicion, which enhanced my analysis efficiency.''}
The feedback from the experts supports the usefulness of \textit{FraudAuditor} to help discover patterns in the data efficiently without having to directly query and interpret the boring raw data.

\textbf{System usability.} Experts agree that the proposed visual analytics approach follows a real health insurance audit workflow.
From overviews to details, each step is easy to understand (ratings: 4.3/5).
\textit{``Compared to the traditional audit process, the system can greatly shorten the time of manual analysis, and the information provided by the system is what I need in the analysis,''} E1 remarked.
E6 commented: \textit{``the system delivers user-friendly interactions, and the analysis process is reasonable.''}
\rc{Evaluations on the learning cost were also presented.}
E5 stated, \textit{``When I first started using this system, I had to read the documentation frequently. The system has a different interface than the health insurance database management system. It is recommended that the visual analysis system guide users during usage.''}
E5 added, \textit{``I want to use the system for my work. Once I am familiar with the process, I can better assess the effectiveness of fraud detection algorithms.''}
Based on experts' feedback, \textit{FraudAuditor} can assist experts in detecting, analyzing, and labeling fraudulent groups.

\rc{\textbf{Suggestions.} Experts also offered valuable suggestions for \textit{FraudAuditor}. E3 suggested that the system could integrate more details of tests, examinations, and procedures in healthcare (currently not available in our dataset), which would help improve the accuracy of fraud verification. E1 also requested additional granular data filtering options, such as selecting data from a particular pharmacy, because sometimes they receive reports of fraud cases and then need to independently extract the relevant data for review. E2 added, \textit{``if the system could support real-time fraud detection and make timely warnings of fraud, it would be more effective in reducing the loss of health insurance funds.''}}

\section{Discussion}\label{sec:discussion}

This section discusses the advantages and limitations of \textit{FraudAuditor} from the perspectives of generalizability and scalability. We also summarize the lessons learned from the implementation process and shed light on future work.

\textbf{Generalizability.} 
\rc{FraudAuditor is designed to detect collusive fraud in healthcare, where patients often visit certain medical institutes together at similar times. By adjusting the co-visit definition, the detection approach can uncover other types of healthcare fraud. For example, in the case where fraudsters work asynchronously, co-visit behaviors can be detected by increasing the parameter of the co-visit time gap or using time alignment methods (e.g., DTW~\cite{muller2007dynamic}); in the case where fraudsters spatially disperse, the location constraint can be relaxed to the same area by considering the geographic information of the medical institutes.}

Our approach can also be generalized to other application domains where the frauds share similar characteristics of group and simultaneity, e.g., electronic commerce fraud, spam detection, and telecommunication fraud. Taking spam as an example, spammers always use botnets to send group emails by controlling multiple bot accounts. Similar to the co-visit network in health insurance, a co-sending network between email accounts can be constructed by considering the sending interval and the number of co-sendings. 

\textbf{Scalability.} 
\rc{Due to enormous volumes of raw data, our system faces scalability issues in data processing and data visualization. We allow users to use data filtering to focus on a small set of records. Since the percentage of fraud is small, reducing the data scope through spatio-temporal segmentation and attribute filtering, such as removing visit records from highly regulated public hospitals, is in line with the practical workflow and the principle of overview to detail in visualization~\cite{shneiderman2003eyes}. In the extreme case of a group with many fraudsters, the patient behavior view may encounter visual clutter and rendering bottlenecks, which can be mitigated using data sampling and progressive visualization~\cite{zgraggen2016progressive}.}


\textbf{Lessons learned.} 
First, multi-level views should be provided for the visualization of complex high-dimensional data to support progressive analysis.
Directly presenting all information to users increases cognitive load.
The health insurance data used in this paper involves associations among multiple subjects and has a large number of attribute dimensions.
We split the analysis tasks and coordinated views into three levels (i.e., overview-level, group-level, and patient-level).
Thus, both the visual analytics approach and the system design follow the overview-to-detial principle.

Second, intuitive and effective visualization helps users learn quickly.
Because users of our system are experts with a background in health insurance audit, they don't know much about visualization.
The charts in \textit{FraudAuditor} are mainly common and popular charts, such as bar charts and node-link diagrams, which help lower the learning cost for users and increase their trust in the system.

\textbf{Future work.} In the future, \rc{to detect other types of fraud, such as doctor-patient collusion}, we plan to provide more detailed contexts of medical records by constructing a dynamic heterogeneous network of patients, doctors, and medical institutions. Another possible direction is to reduce the cost of manually labeling the dataset by leveraging active learning techniques to improve the efficiency of data instance selection. Additionally, the precision of group detection can be further improved by semi-supervised algorithms. \rc{We also plan to add more guidance and annotations to the system to further improve its usability.}

\section{Conclusion}\label{sec:ccl}

In this paper, we proposed a visual analytics approach that supports the identification, examination, and annotation of collusive fraud in health insurance. The design and implementation of \textit{FraudAuditor} are based on close collaboration with domain experts. By leveraging both automated algorithms and human experience, \textit{FraudAuditor} supports a multi-level fraud analysis, including the co-visit network overview, suspicious groups identification, and suspicious patients examination. A suite of visualization designs supports the detection and exploration of fraud groups. The effectiveness of our approach and the usability of the prototype system were recognized through case studies and interviews involving health insurance audit experts.


%



\ifCLASSOPTIONcompsoc
  \section*{Acknowledgments}
\else
  \section*{Acknowledgment}
\fi

This work was supported by the NSFC (62132017, 62202244).

\ifCLASSOPTIONcaptionsoff
  \newpage
\fi



\bibliographystyle{IEEEtran}
\bibliography{template}
%



%


\begin{IEEEbiography}[{\includegraphics[width=1in,height=1.25in,clip,keepaspectratio]{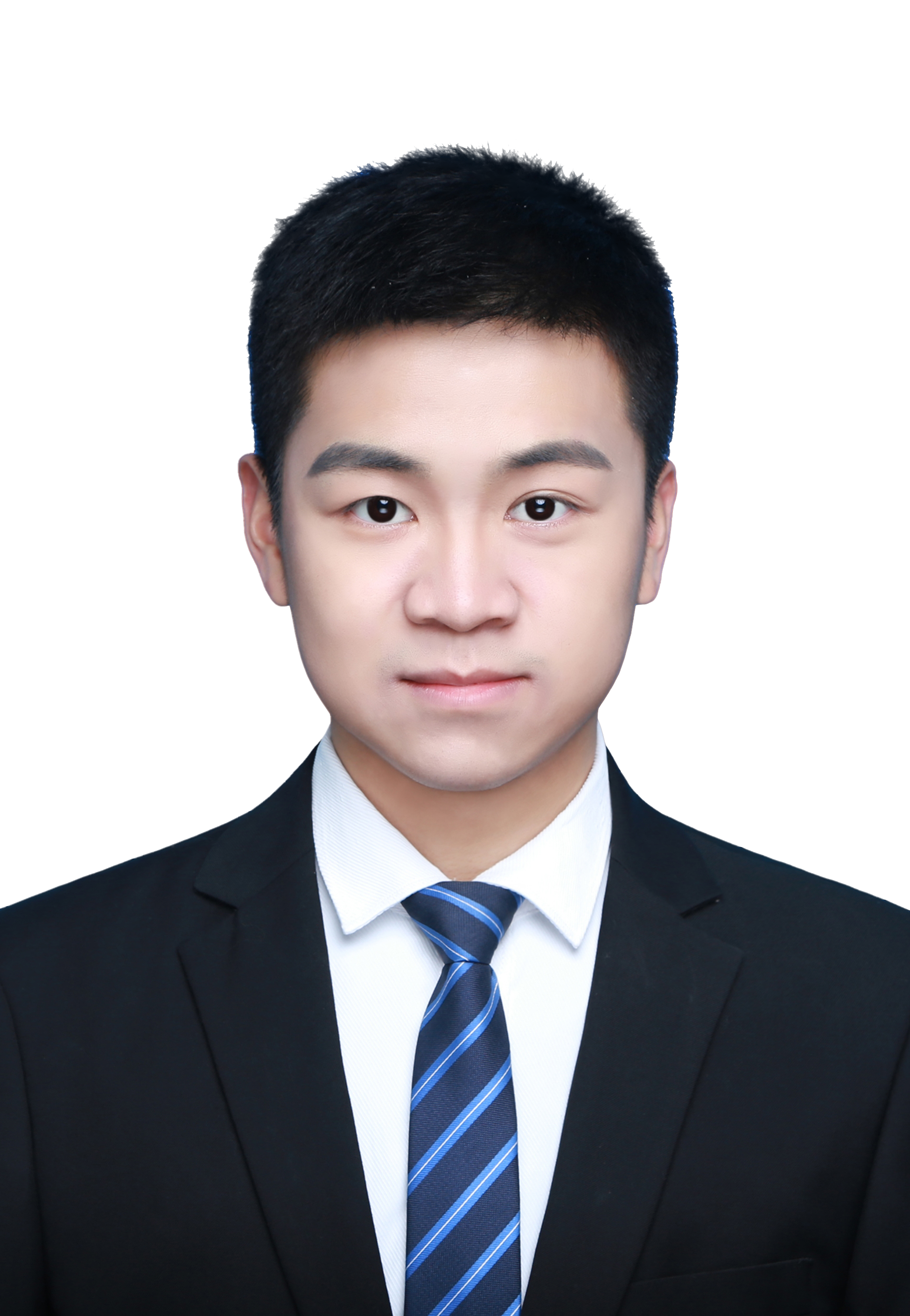}}]{Jiehui Zhou}
is a Ph.D. candidate at the State Key Laboratory of CAD\&CG, Zhejiang University, and works under the supervision of Prof. Wei Chen. He holds a bachelor's degree in computer science and technology from Central South University. His main research focuses on visual analytics, decision intelligence, and human-centered AI, with a special interest in their applications in healthcare.
\end{IEEEbiography}

\begin{IEEEbiography}[{\includegraphics[width=1in,height=1.25in,clip,keepaspectratio]{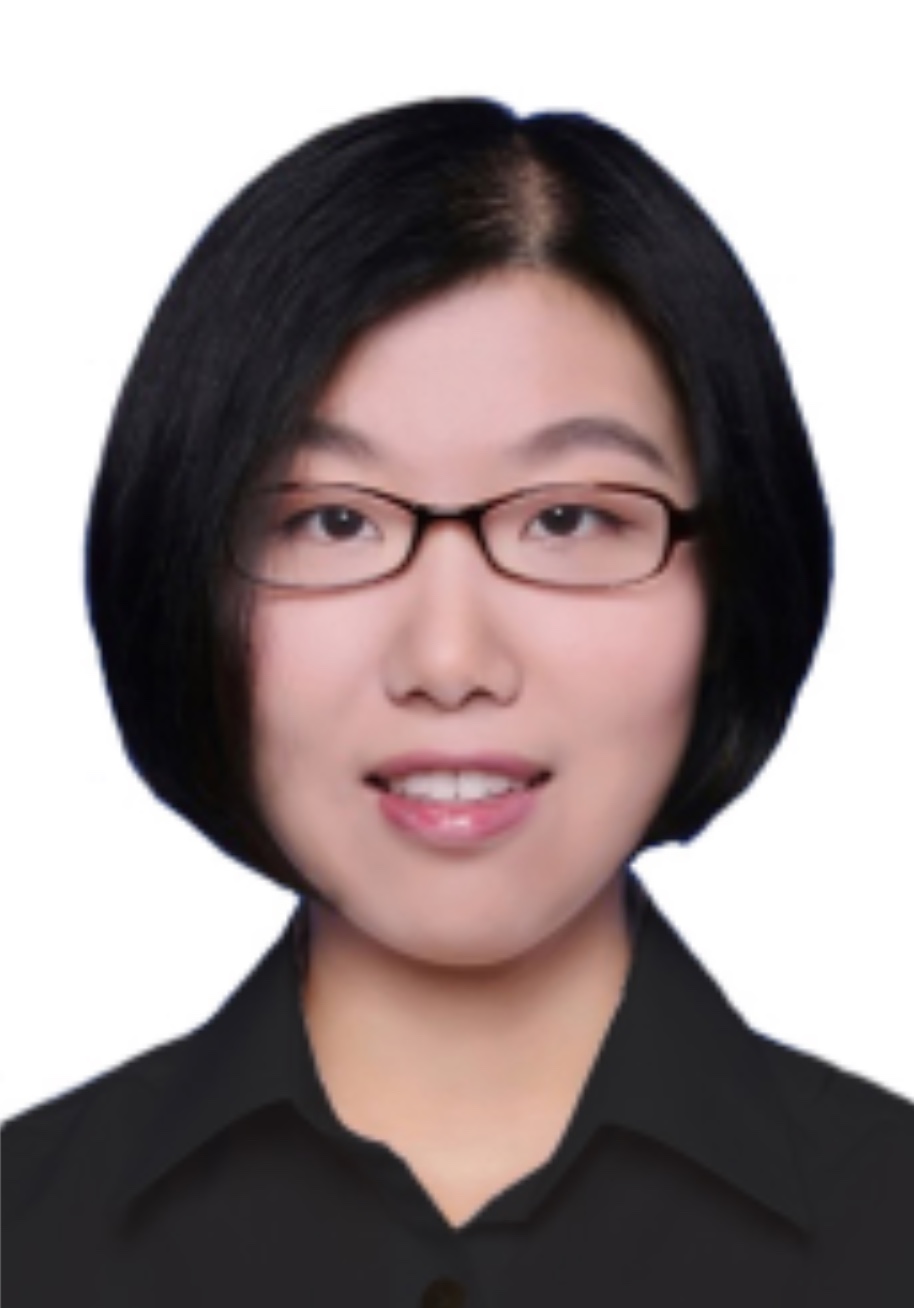}}]{Xumeng Wang}
is a lecturer of computer science in Nankai University. She received the Ph.D. degree in computer science and technology from Zhejiang University in 2021. Her research interests are visual analytics and privacy preservation.
\end{IEEEbiography}

\begin{IEEEbiography}[{\includegraphics[width=1in,height=1.25in,clip,keepaspectratio]{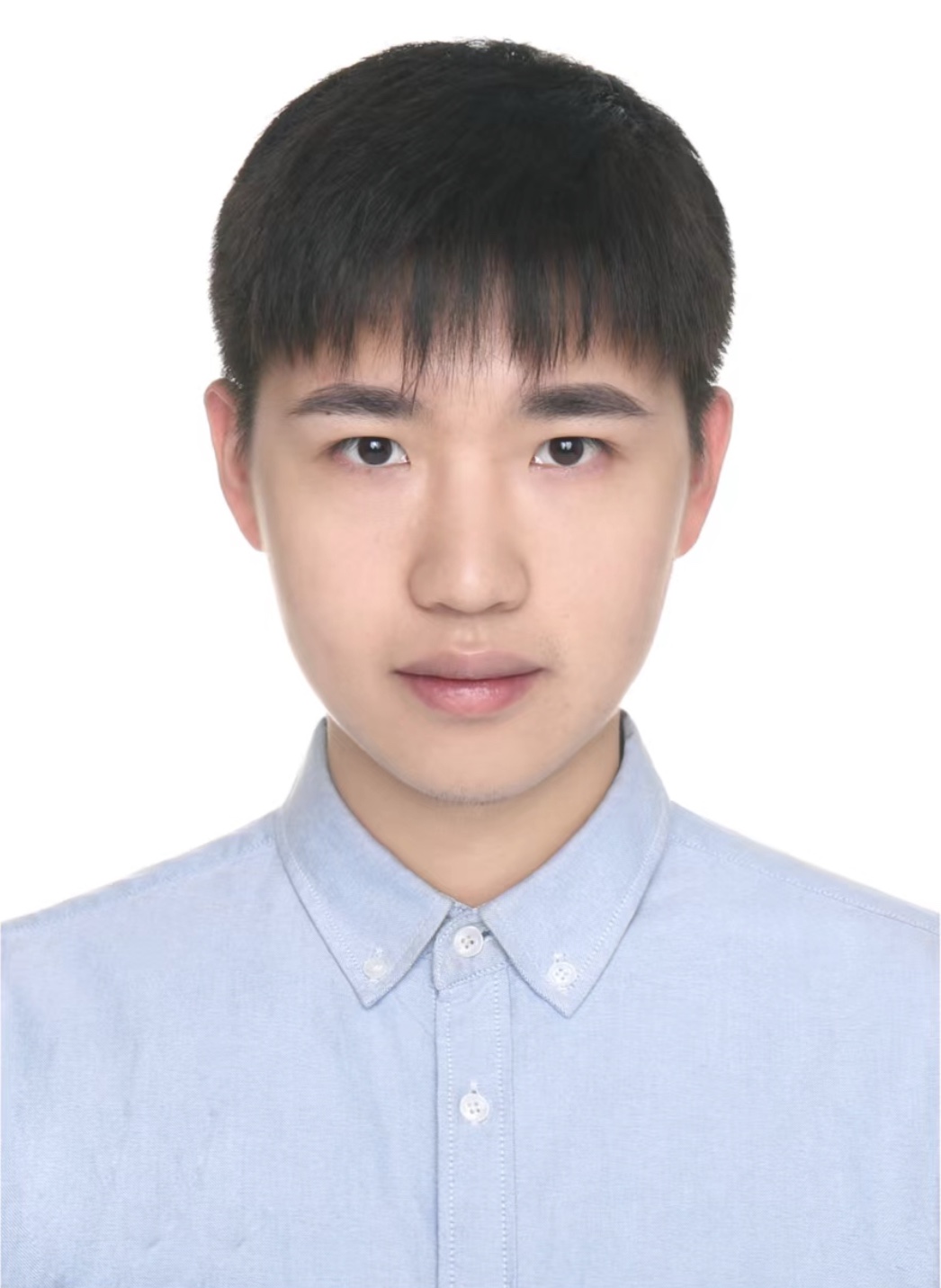}}]{Jie Wang}
is a development engineer for business intelligence at Alibaba Group, Hangzhou. He earned a M.S. degree in software engineering from Zhejiang University in 2021. His research focuses on information visualization and augmented analytics.
\end{IEEEbiography}

\begin{IEEEbiography}[{\includegraphics[width=1in,height=1.25in,clip,keepaspectratio]{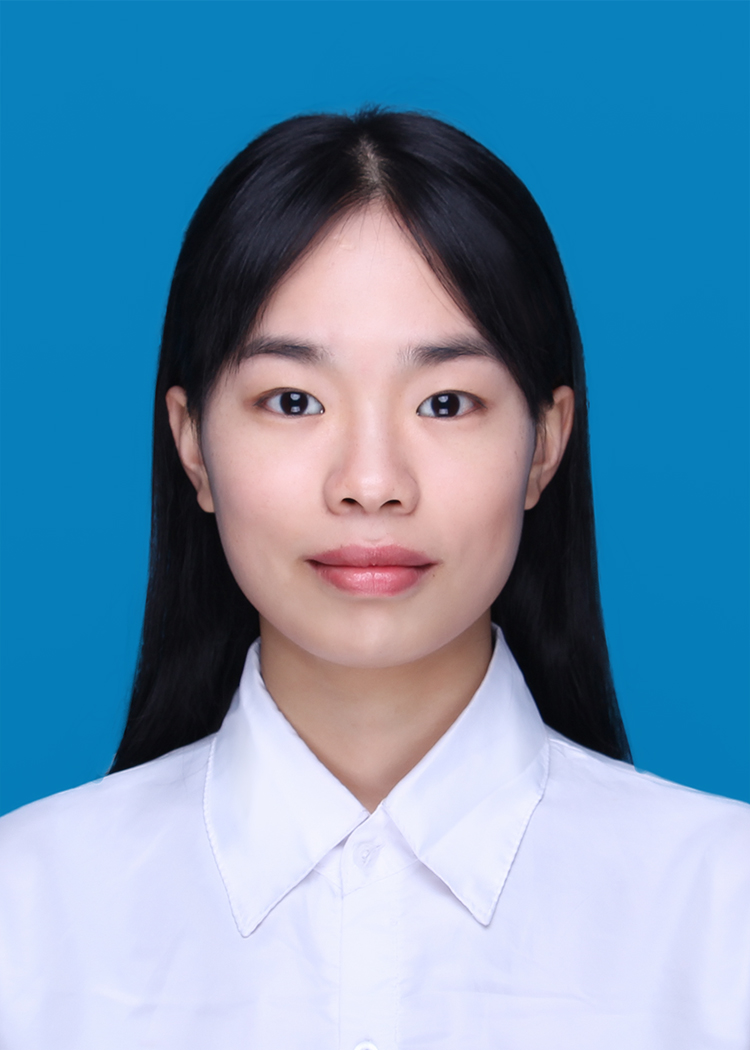}}]{Hui Ye}
is a development engineer at Tencent, Shenzhen. She earned a M.S. degree in software engineering from Zhejiang University in 2021. Her research focuses on data visualization.

\end{IEEEbiography}

\begin{IEEEbiography}[{\includegraphics[width=1in,height=1.25in,clip,keepaspectratio]{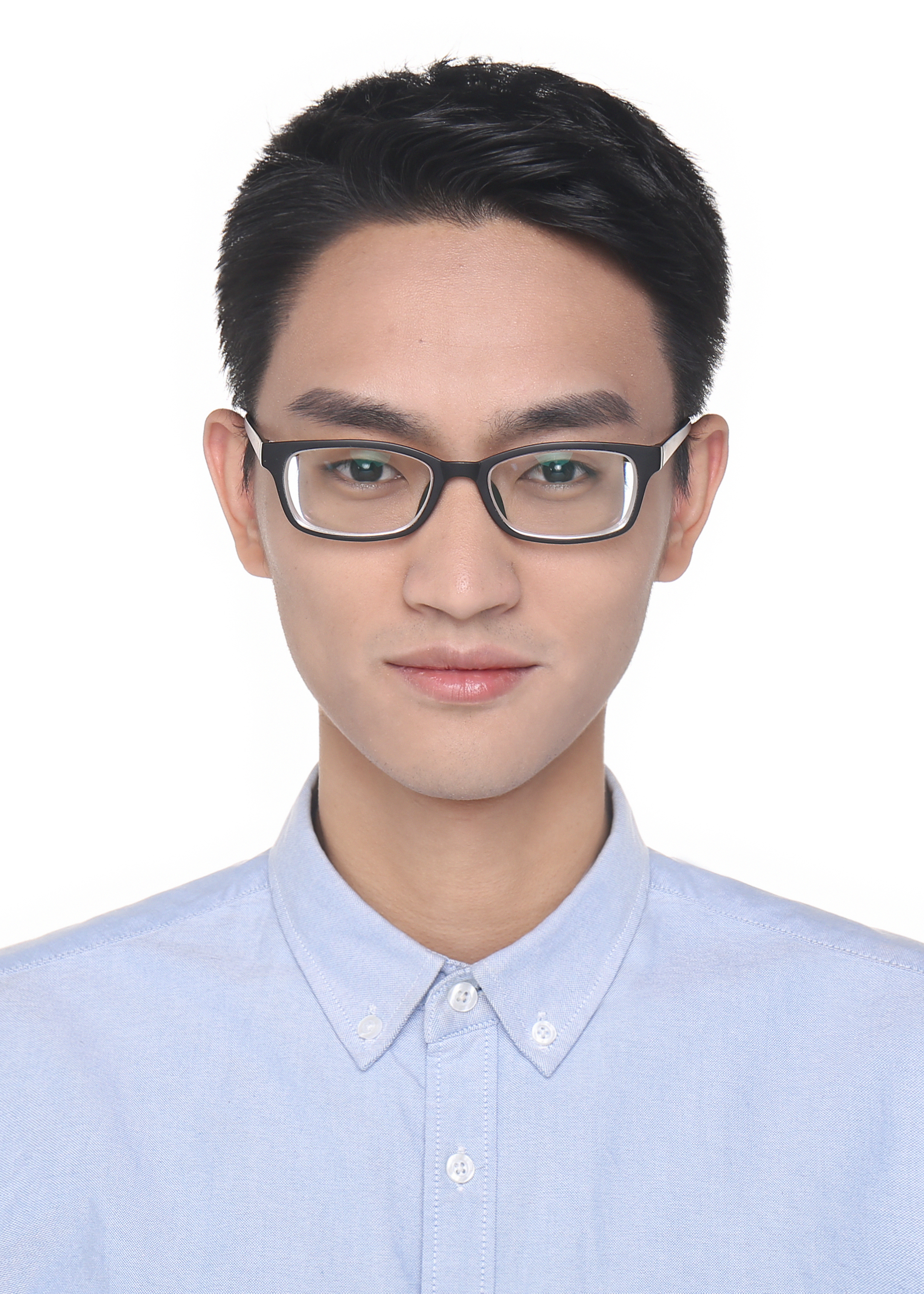}}]{Huanliang Wang}
is currently working toward the M.S. degree with the State Key Lab of CAD\&CG, Zhejiang University, Hangzhou, China. His research interests are data visualization and visual analytics.
\end{IEEEbiography}


\begin{IEEEbiography}[{\includegraphics[width=1in,height=1.25in,clip,keepaspectratio]{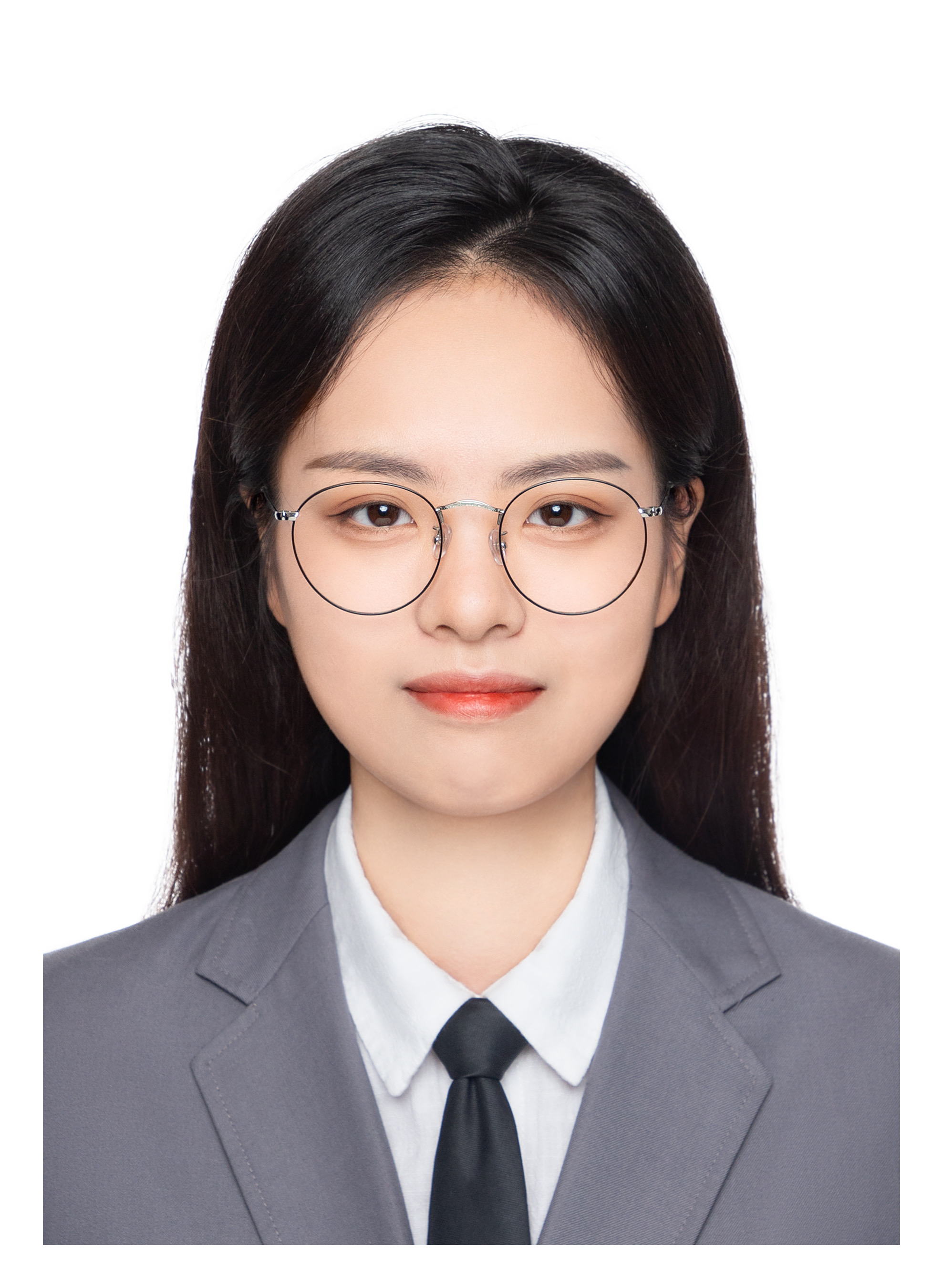}}]{Zihan Zhou}
is a Ph.D. candidate in the State Key Lab of CAD\&CG at Zhejiang University, Hangzhou. She earned a B.S. degree in digital media technology from Zhejiang University in 2021. Her research interests are data visualization and visual analytics.
\end{IEEEbiography}

\begin{IEEEbiography}[{\includegraphics[width=1in,height=1.25in,clip,keepaspectratio]{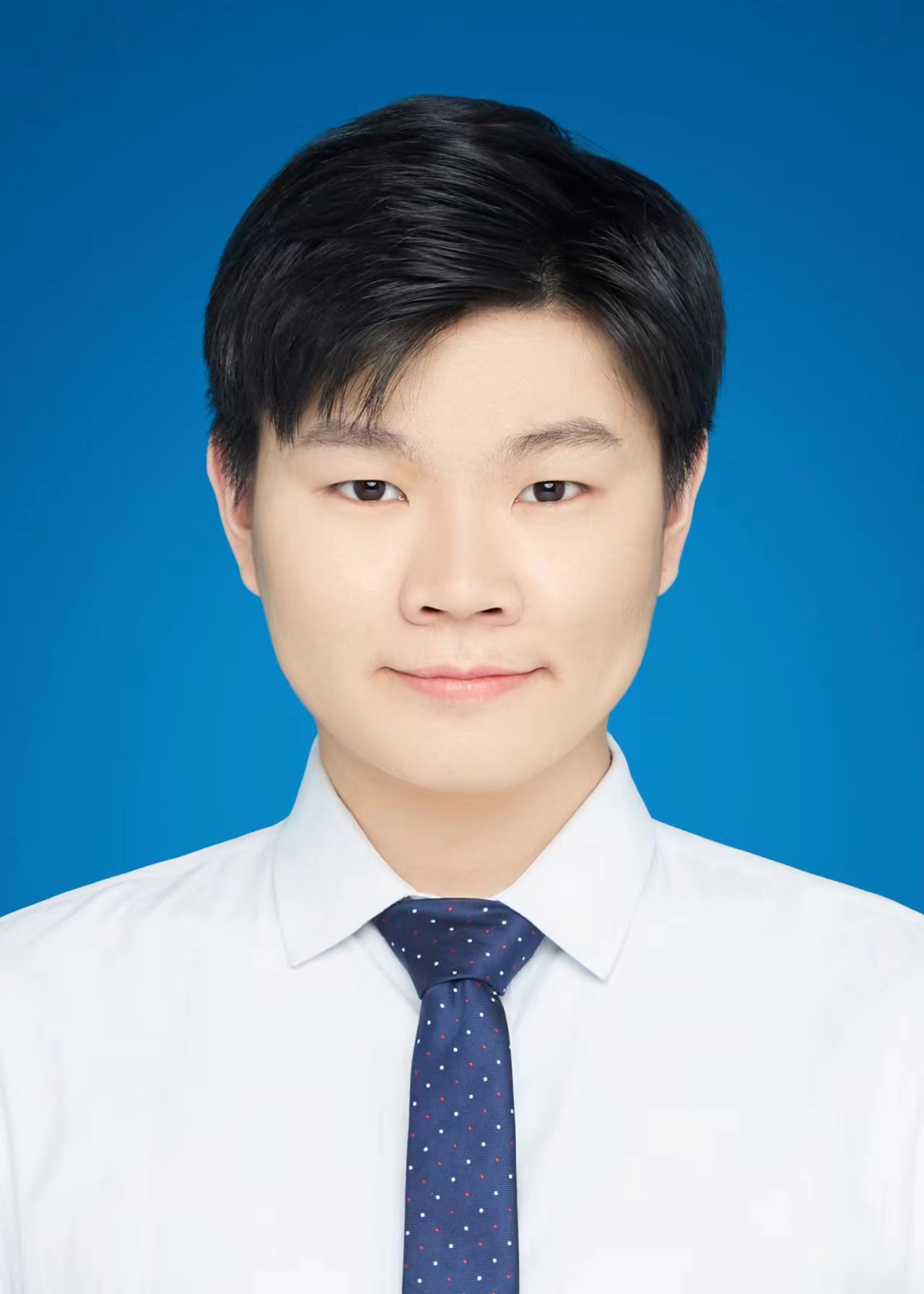}}]{Dongming Han}
is currently working toward the Ph.D. degree with the State Key Lab of CAD\&CG, Zhejiang University, Hangzhou, China. His research interests include information visualization, graph visualization, and visual analytics. He received a B.S. degree in software engineering from Zhejiang University in 2017.
\end{IEEEbiography}

\begin{IEEEbiography}[{\includegraphics[width=1in,height=1.25in,clip,keepaspectratio]{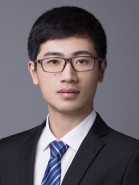}}]{Haochao Ying}
is currently an assistant professor in the School of Public Health, Zhejiang University. He received the Ph.D. degree in the College of Computer Science from Zhejiang University in 2019, and the B.S. degree in computer science and technology from Zhejiang University of Technology in 2014. His research interests include data mining for healthcare and personalized recommender systems. He has authored some papers at prestigious international conferences and journals, such as TKDE, TCBB, JBHI, IJCAI, ICML, and CVPR.
\end{IEEEbiography}

\begin{IEEEbiography}[{\includegraphics[width=1in,height=1.25in,clip,keepaspectratio]{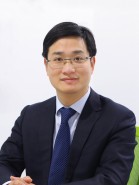}}]{Jian Wu}
is a full professor at Zhejiang University. He is currently the director of the Real Doctor AI Research Centre of Zhejiang University. His research interests include artificial intelligence, data mining, and their applications in healthcare and biomedicine. He received his Ph.D. degree in Computer Science and Technology from Zhejiang University. He has published more than 200 papers in some prestigious refereed journals and conference proceedings, such as IEEE Transactions on Knowledge and Data Engineering, IEEE Transactions on Medical Imaging, CVPR, IJCAI, AAAI, ICML, and MICCAI. He is a distinguished member of the CCF.
\end{IEEEbiography}

\begin{IEEEbiography}[{\includegraphics[width=1in,height=1.25in,clip,keepaspectratio]{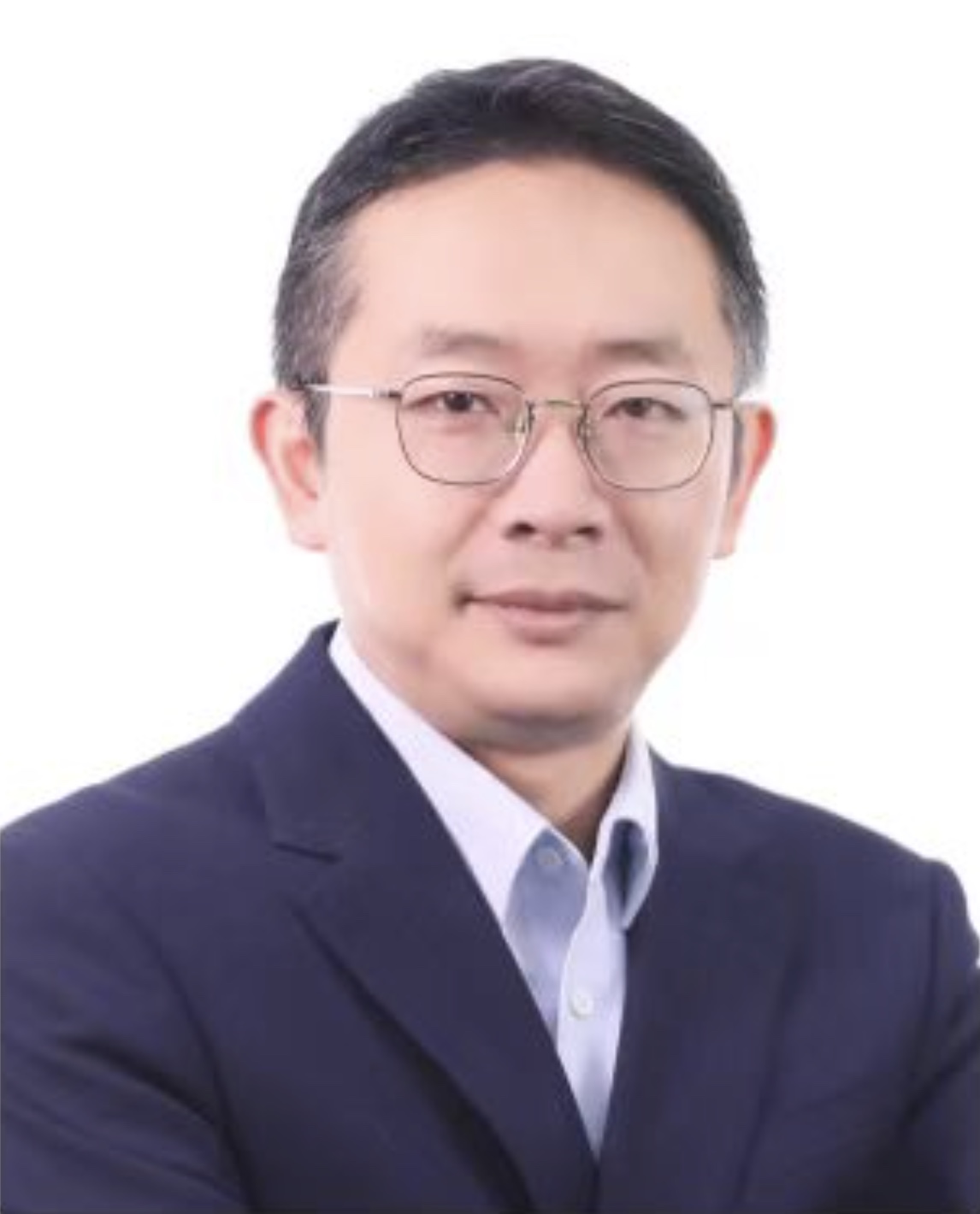}}]{Wei Chen} is a professor at the State Key Lab of CAD\&CG, Zhejiang University. His research interests include visualization and visual analytics. He has published more than 90
IEEE/ACM Transactions and IEEE VIS papers. He actively served as a guest or associate editor of
IEEE Transactions on Visualization and Computer Graphics, IEEE Transactions on Intelligent Transportation Systems, and Journal of Visualization. For more information, please refer to
\url{http://www.cad.zju.edu.cn/home/chenwei/}
\end{IEEEbiography}



\end{document}